\documentclass[12pt]{spieman}  
\usepackage[]{graphicx}
\usepackage{setspace}
\usepackage{tocloft}
\usepackage{amssymb}

\title{The Speedster-EXD- A New Event-Driven Hybrid CMOS X-ray Detector} 

\author{Christopher V. Griffith\supscr{a}, Abraham D. Falcone\supscr{a}, Zachary R. Prieskorn\supscr{a}, David N. Burrows\supscr{a}}

\affiliation{\supscrsm{a}The Pennsylvania State University, Astronomy \& Astrophysics, 525 Davey Lab, University Park, USA, 16802\\
}

\cftpagenumbersoff{figure}
\cftpagenumbersoff{table} 
\begin{document} 
\maketitle 

\begin{abstract}
The Speedster-EXD is a new 64x64 pixel, 40 $\mu$m pixel pitch, 100 $\mu$m depletion depth hybrid CMOS X-ray detector (HCD) with the capability of reading out only those pixels containing event charge, thus enabling fast effective frame rates. A global charge threshold can be specified, and pixels containing charge above this threshold are flagged and read out.  The Speedster detector has also been designed with other advanced in-pixel features to improve performance, including a low-noise, high-gain CTIA amplifier that eliminates interpixel capacitance crosstalk (IPC), and in-pixel Correlated Double Sampling (CDS) subtraction to reduce reset noise.  We measure the best energy resolution on the Speedster-EXD detector to be 206 eV (3.5 \%) at 5.89 keV and 172 eV (10.0 \%) at 1.49 keV.  The average IPC to the four adjacent pixels is measured to be 0.25 $\pm$ 0.2 \% (i.e. consistent with zero).  The pixel-to-pixel gain variation is measured to be 0.80 $\pm$ 0.03 \%, and a Monte Carlo simulation is applied to better characterize the contributions to the energy resolution.  



\end{abstract}


{\noindent \footnotesize{\bf Address all correspondence to}: Christopher V. Griffith, The Pennsylvania State University, Astronomy \& Astrophysics, 525 Davey Lab, University Park, USA, 16802; E-mail:  \linkable{cvg5124@psu.edu}. 

\noindent Abe Falcone,   The Pennsylvania State University, Astronomy \& Astrophysics, 525 Davey Lab, University Park, USA, 16802; E-mail:  \linkable{adf15@psu.edu}.}

\begin{spacing}{1}   

\section{Introduction}
\label{sect:intro}  
Current X-ray space missions, such as Chandra and XMM-Newton, have produced exceptional science results using Charge-Coupled Devices (CCDs) in their detector focal planes. CCDs currently provide Fano-limited energy resolution ($\Delta$E/E $\sim$ 2.0 \% at 5.89 keV) with $\lesssim$ 2 e$^{-}$ read noise \cite{2003SPIE.4851...28G}.   
However, X-ray missions on the horizon, such as SMART-X \cite{2012SPIE.8443E..16V} and/or the X-ray Surveyor \cite{2014arXiv1401.3741K}, are slated to have larger collecting areas up to 30 times that of current missions, which will increase the X-ray photon rate from even moderate to low flux sources beyond the frame rate of the fastest CCDs.  These future high-throughput missions have exciting science goals, which include observing gamma ray bursts in the early universe ($z>7$) to explore cosmic structure, observing bright objects such as blazars to detect faint absorption lines in the Warm Hot Ionized Medium (WHIM), and studying supermassive black holes and their environments up to $z>6$.   However, current CCD technology will not be adequate for the focal plane requirements of these future high-throughput missions.

We present the characterization of a new hybrid CMOS X-ray detector (HCD) called the Speedster-EXD, which offers the fast effective frame rates needed for future high-throughput missions.  HCDs are also more radiation hard and have lower power requirements than current CCDs.  The Speedster detector has new in-pixel features that improve upon previous generation HAWAII HCDs\cite{2003SPIE.4850..867L, 2013NIMPA.717...83P, 2012SPIE.8453E..0FG}, increasing the effective frame rate and performance.  We have developed the Speedster to mitigate interpixel capacitance crosstalk (IPC) seen on HAWAII HCDs, and to utilize the in-pixel circuitry capabilities of HCD technology by embedding event-driven circuitry.  In this paper, we briefly discuss HCDs, including their advantages (Section \ref{HCD}) and characteristics (Section \ref{chara_astronomy}), followed by the description of the Speedster-EXD detector and its new features (Section \ref{sect:speedster}). We then discuss our testing of the Speedster-EXD detector (Sections 5-7), including its measured IPC (Section \ref{sect:ipc}), read noise (Section \ref{read_noise}), energy resolution (Section \ref{energy_res}), dark current (Section \ref{sect:dark_current}), and gain variation (Section \ref{gain_var}).

\section{Hybrid CMOS Detectors}
\label{HCD}
\begin{figure}
   \begin{center}
   \begin{tabular}{c}
   \includegraphics[height=5.5cm]{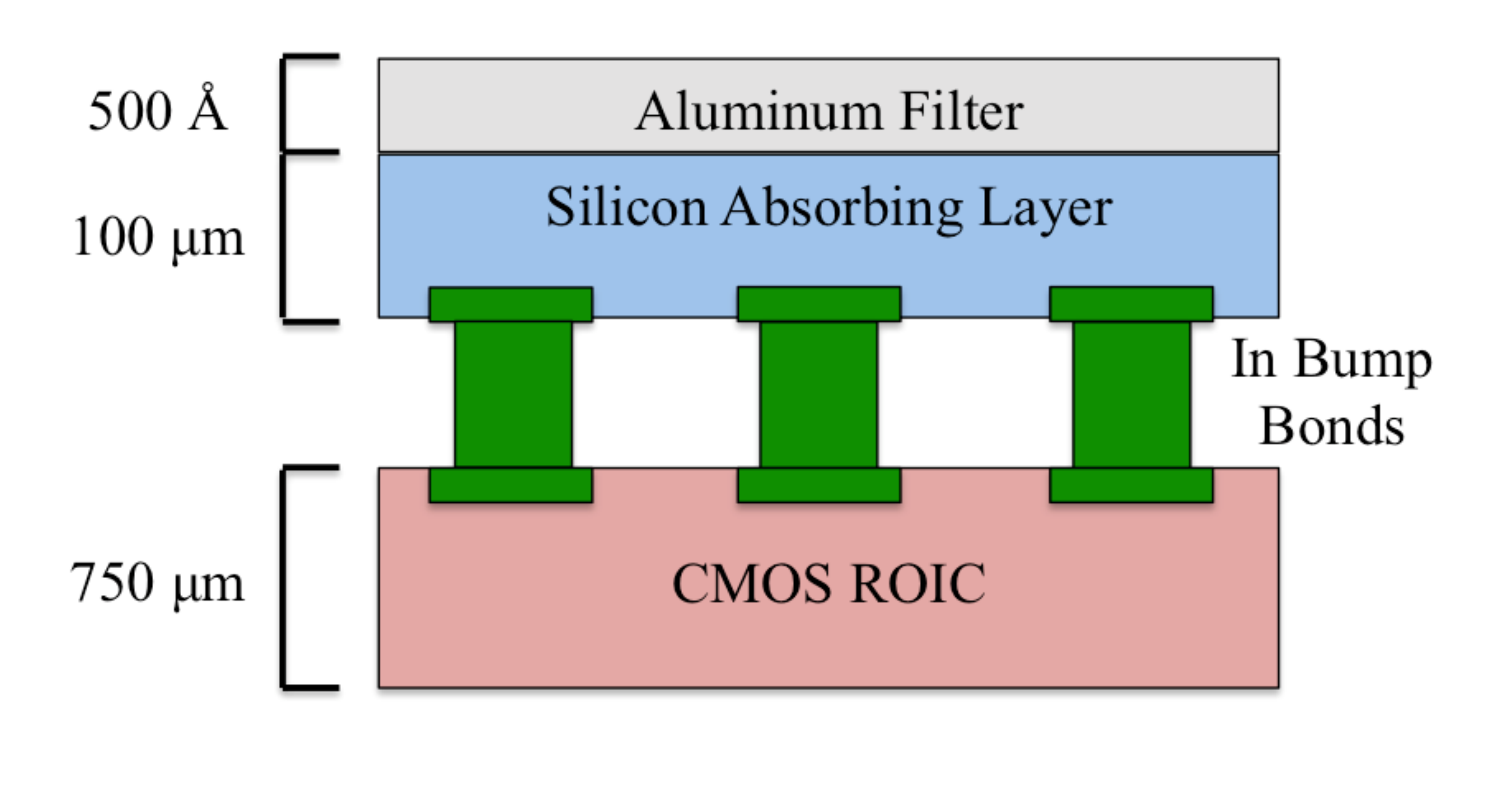}
   \end{tabular}
   \end{center}
   \caption 
   { \label{fig:cmos_cartoon} 
Schematic of hybrid CMOS X-ray detector (not to scale). The dimensions of the aluminum filter, silicon absorbing layer and ROIC layer of the Speedster-EXD detector are shown.  } 
   \end{figure}

Hybrid CMOS detectors (HCDs) consist of two layers; an absorbing layer (silicon or HgCdTe depending on the application) and a readout integrated circuit (ROIC) layer.  The absorbing layer is optimized for high quantum efficiency, and the ROIC layer is optimized for fast readout and advanced in-pixel electronics.  The two layers are then indium bump bonded together at each pixel.  X-ray HCDs use silicon as the absorbing substrate. A thin layer of aluminum is deposited directly on the silicon to block optical light.  A schematic of an X-ray HCD can be seen in Figure \ref{fig:cmos_cartoon}.

The hybrid nature of X-ray HCDs allows for the use of high resistivity silicon and therefore offers high depletion depths ($>$100 $\mu$m).  This high depletion depth gives HCDs high quantum efficiency at energies above 6 keV, and they can be used to detect X-rays from 0.2 - 20 keV \cite{2014SPIE.9154E..10P}.     

The astronomy community has utilized HCD technology in the infrared and optical bands\cite{2008SPIE.7021E..0HB, 2008SPIE.7021E..02B}.  Infrared HCDs, which use a HgCdTe absorbing layer, have already flown on the Hubble Space Telescope\cite{2008SPIE.7010E..1EK} and the Wide-Field Infrared Survey Explorer (WISE) \cite{2010AJ....140.1868W} mission, and they are also slated to fly on the James Webb Space Telescope (JWST)\cite{2014PASP..126..739R}.  Optical HCDs (Hybrid Visible Silicon Imagers, HyViSi) have flown on the Orbiting Carbon Observatory 2 (OCO-2)\cite{2014SPIE.9241E..05B} and the Mars Reconnaissance Orbiter\cite{2007JGRE..112.5S01Z}.  


HCDs offer many advantages over current state-of-the-art CCDs.  The advantages are listed below.  

{\bf Faster Readout Rate:} As mentioned above, CCDs do not have the readout speed required by high-throughput, future X-ray missions.  CCDs will suffer from ``pile-up'', where two or more X-ray photons interact in the same pixel before the frame is read out.  Energy information is lost as the two X-ray photons are detected as a single photon.   CCDs suffer from pile-up on current missions when observing bright objects, and this problem will be amplified with the increased collecting area on future missions \cite{2000ExA....10..439L}.  

HCDs offer much faster readout rates and the ability to read out individual pixels to collect photons from the source of interest before pile-up occurs.  As will be discussed in Section \ref{sect:speedster}, the Speedster-EXD detector has the ability to read out only pixels containing charge above an adjustable threshold, increasing effective frame rates by orders of magnitude over CCDs.  

{\bf Less Susceptible to Radiation Damage:} CCDs are more susceptible to radiation damage than HCDs due to their bucket brigade readout scheme\cite{2005SPIE.5898..201G}.  A CCD's charge transfer efficiency (CTE) degrades over time as the detector is exposed to damaging alpha particle and proton radiation\cite{2012SPIE.8453E..36B,1991ExA.....2..179L}.  The direct readout of each pixel on an HCD reduces this problem as the charge is only transferred through the absorber thickness ($\sim$100 $\mu$m) rather than across centimeters of the detector surface.  This decrease in charge transfer distance makes HCDs much more radiation hard than CCDs ($>$100 krads).
  
{\bf Less Susceptible to Micrometeoriod Damage:} Micrometeoroids have caused serious damage to CCD detectors on current missions, such as XMM-Newton\cite{2001A&A...375L...5S}.  Front-side illuminated CCDs have exposed gates, which makes them vulnerable to catastrophic damage. The bucket brigade readout design on front-side illuminated and back-side illuminated CCDs causes damage to a single pixel to affect all pixels in the same column on the detector.  HCDs do not have exposed gates since all of the electronics are behind the absorbing layer, and the direct readout of pixels confines any damage to only the impacted pixels.  

{\bf Low Power Use:}  Due to the large capacitive loads in the parallel clock gates, the bucket-brigade charge transfer technique on CCDs requires a lot of power.  For example, the Swift XRT instrument uses 8.4 Watts to clock the signal through the X-ray CCD.  HCDs have much smaller capacitive loads that involve low power CMOS switches \cite{2002SPIE.4836..247K}.  A typical 1024 x 1024 pixel$^{2}$ H1RG HCD, which uses a source follower amplifier in each pixel, has been shown to use $\sim$200 mW for all bias clock generation and readout\cite{2012SPIE.8453E..0EF}.  The CTIA amplifier and in-pixel comparator used in the Speedster-EXD pixel design (see Section \ref{sect:speedster}) use more power per pixel than a simple source follower amplifier.  A conservative calculation gives an estimated power use of $\sim$1.6 W for a 1024 x 1024 pixel$^{2}$ Speedster detector.  The low power use of both pixel designs reduces the overall cost of a mission and enables large format arrays to be used with a low power budget.  


\section{Characterizing Hybrid CMOS Detectors for X-ray Astronomy} \label{chara_astronomy}
The parameters tested on the Speedster-EXD detectors are shown below.  Each parameter is described, and its importance to the development of HCD technology is discussed. 

\subsection{Interpixel Capacitance Crosstalk} \label{ipc_exp}
\begin{figure}
   \begin{center}
   \begin{tabular}{c}
   \includegraphics[height=5.5cm]{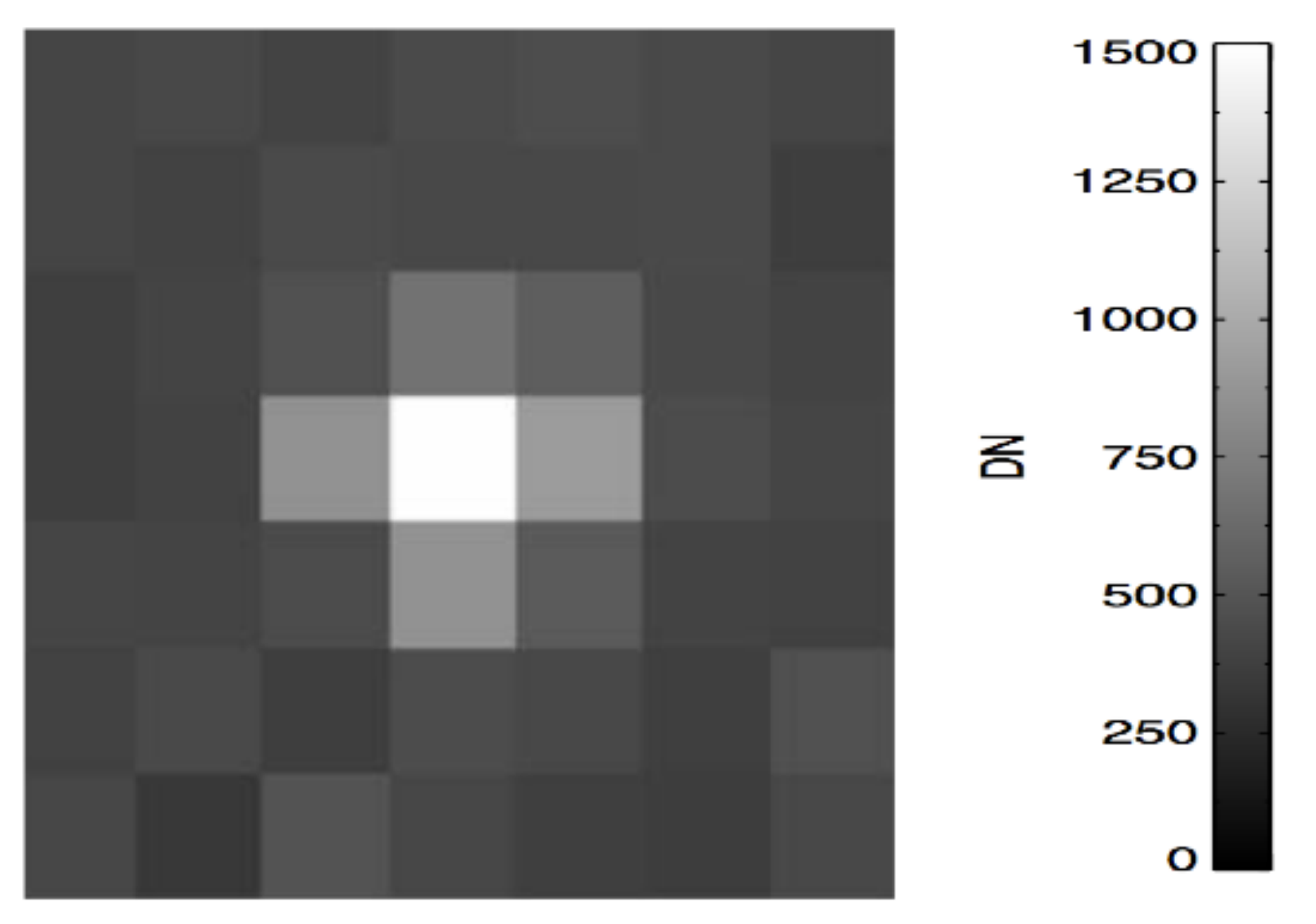}
   \end{tabular}
   \end{center}
   \caption 
   { \label{fig:ipc_event} 
An example X-ray event with high IPC.  Color bar shows brightness as a function of DN (Digital Number). } 
   \end{figure}

IPC is unwanted signal spreading between neighboring pixels.  It is caused by unintended, parasitic capacitances between adjacent pixels inside the silicon absorber array \cite{2006SPIE.6276E..0FF}.  We have previously measured IPC on HAWAII HCDs\cite{2013NIMPA.717...83P}.  An example X-ray event with high IPC is shown in Figure \ref{fig:ipc_event}.  Signal is coupled into the up, down, left, and right pixels, making it difficult to properly grade the events, and distinguish between actual charge spreading and signal coupling \cite{2005SSRv..120..165B}.  To a lesser degree, signal can also be lost to pixels outside the 3x3 region surrounding the central pixel.  IPC degrades the energy resolution of an event since more pixels must be included in a measurement to fully contain the total signal, thus increasing the total noise.  Also, proper event grading cannot be achieved for all events when a detector has large IPC.  

\subsection{Read Noise}
Read noise is associated with the conversion of the signal charge generated by absorption of an X-ray into an electronic voltage signal within the ROIC.  Read noise sets the noise floor of a detector, thereby determining the lower limit on energy resolution.  Some recent HCD developments are aimed at lowering read noise to enable missions to capitalize on HCD advantages without sacrificing the excellent noise performance of CCDs.

\subsection{Energy Resolution}
Energy resolution is an important factor for an X-ray detector and determines how well spectral lines can be resolved in observations.  CCDs currently have Fano-limited \cite{1947PhRv...72...26F} energy resolution ($\Delta$E/E$\sim$2.0 \% at 5.89 keV).  The energy resolution of HAWAII HCDs for X-ray detection has been dominated by IPC and high read noise \cite{2013NIMPA.717...83P, 2012SPIE.8453E..0FG}.

\subsection{Dark Current}
Dark current arises from thermal energy creating unwanted electron-hole pairs in the silicon substrate.  HCDs must be cooled in order to reduce the dark current.  Dark current is measured on the Speedster-EXD detectors as a function of temperature.

\subsection{Gain Variation}
In order to accomplish the individual readout of each pixel on an HCD, the ROIC bonded to each pixel contains an amplifier.  The gain of these amplifiers can vary slightly across the detector.  These pixel-to-pixel variations can cause differences in energy measurement of X-ray photons as a function of where they impact the absorber.  This difference in energy measurements causes the energy resolution to be degraded, unless corrections are made to account for the gain variations.  We measure the magnitude of the gain variations across the HCDs, and we model its effect on the measured energy resolution.

\section{Speedster-EXD Detector}
\label{sect:speedster} 
\begin{figure}
   \begin{center}
   \begin{tabular}{c}
   \includegraphics[height=6.5cm]{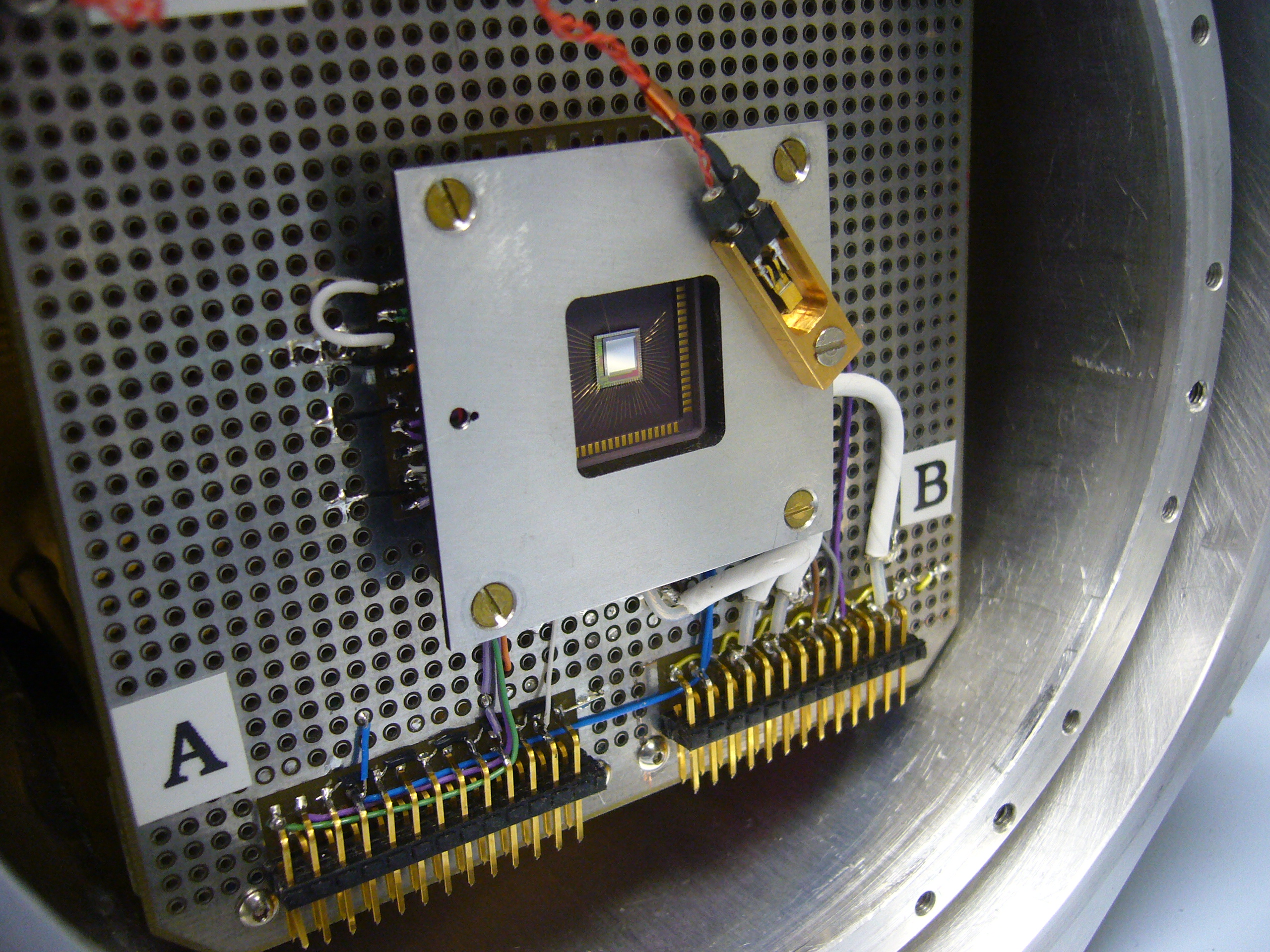}
   \end{tabular}
   \end{center}
   \caption 
   { \label{fig:speedster_pic} 
Speedster-EXD detector located on a dewar breadboard placed inside a test dewar.  The dewar bread board shown is 10 cm x 10 cm.    } 
   \end{figure}

The Speedster-EXD is a 64x64 pixel$^{2}$ prototype Si X-ray HCD with a pixel pitch of 40 $\mu$m and depletion depth of 100 $\mu$m.  It was designed as a part of a collaboration between Penn State University and Teledyne Imaging Systems (TIS) and fabricated by TIS.  Figure \ref{fig:speedster_pic} shows one of the Speedster-EXD detectors located inside a test dewar.  

The Speedster-EXD detectors have new features that improve upon HAWAII HCDs.  These features are listed below:  


{\bf CTIA Amplifier:}  The Speedster-EXD detector uses a high-gain, low-noise Capacitive Transimpedance Amplifier (CTIA) in each pixel.  HAWAII HCDs use a source follower amplifier, which when combined with the capacitively coupling between pixels, can be the cause of high levels of crosstalk (IPC).    As charge is integrated on the source follower amplifier, the input gate voltage changes, and this voltage change is capacitively coupled to neighboring pixels. Conversely, the CTIA amplifier input gate is held at a constant voltage during integration and is operationally insensitive to IPC.  Therefore, signal cannot be coupled to neighboring pixels, and IPC is eliminated.  The IPC measurement of the Speedster-EXD detector will be discussed further in Section \ref{sect:ipc}.  

{\bf Four Gain Modes:}  The Speedster-EXD has four gain modes to optimize either full well capacity or energy resolution.  Full well capacity is the amount of charge a pixel can receive before saturating.  These four gain modes range from $\sim$0.6 e$^{-}$/DN to $\sim$2.7 e$^{-}$/DN. The low gain modes offer a larger full-well (i.e. the dynamic range is increased) as more X-ray photons can be detected before the full-well capacity  of the pixel is reached.  The high gain modes offer better energy resolution as each detected X-ray photon is converted to a high Digital Number (DN) value.  Depending on the X-ray source of interest, an observer can optimize the gain mode to accomplish their science goal.  



{\bf In-pixel Correlated Double Sampling (CDS) Subtraction:}  The Speedster-EXD has taken full advantage of the in-pixel circuitry capabilities of hybrid CMOS by incorporating CDS subtraction circuitry in each pixel.  Directly after resetting, there is a variable baseline voltage level associated with reset.  CDS enables the subtraction of this variable signal prior to signal integration.  The subtraction is done on-chip, before the signal is amplified.

{\bf Sparse Readout:}  The Speedster-EXD detector also has a new feature called Sparse Readout mode, which provides the ability to read out only pixels that contain event charge.  Each pixel contains a comparator that compares the detected signal in the pixel to an adjustable global threshold.  If the charge detected in the pixel is greater than the set threshold, the pixel is read out.  The position of each row containing one or more pixels with charge that exceeds the set threshold is read into memory.  The columns of the charge containing pixels are found and the pixels are flagged for readout.  Some dead time occurs between each frame ($\sim$18 - 30 microseconds per frame clock cycle, or 1.8 - 3 \% dead time for 1 ms time resolution).  All comparators on the detector are autozeroed before each exposure to avoid any non-uniformity between them.  

Sparse Readout can be run in two modes; single pixel readout, where only the pixels with charge above the set threshold are read out, and 3x3 readout, where the pixels with charge above the set threshold are read out along with the 3x3 pixel region surrounding them.  

\begin{figure}
   \begin{center}
   \begin{tabular}{c}
   \includegraphics[height=7cm]{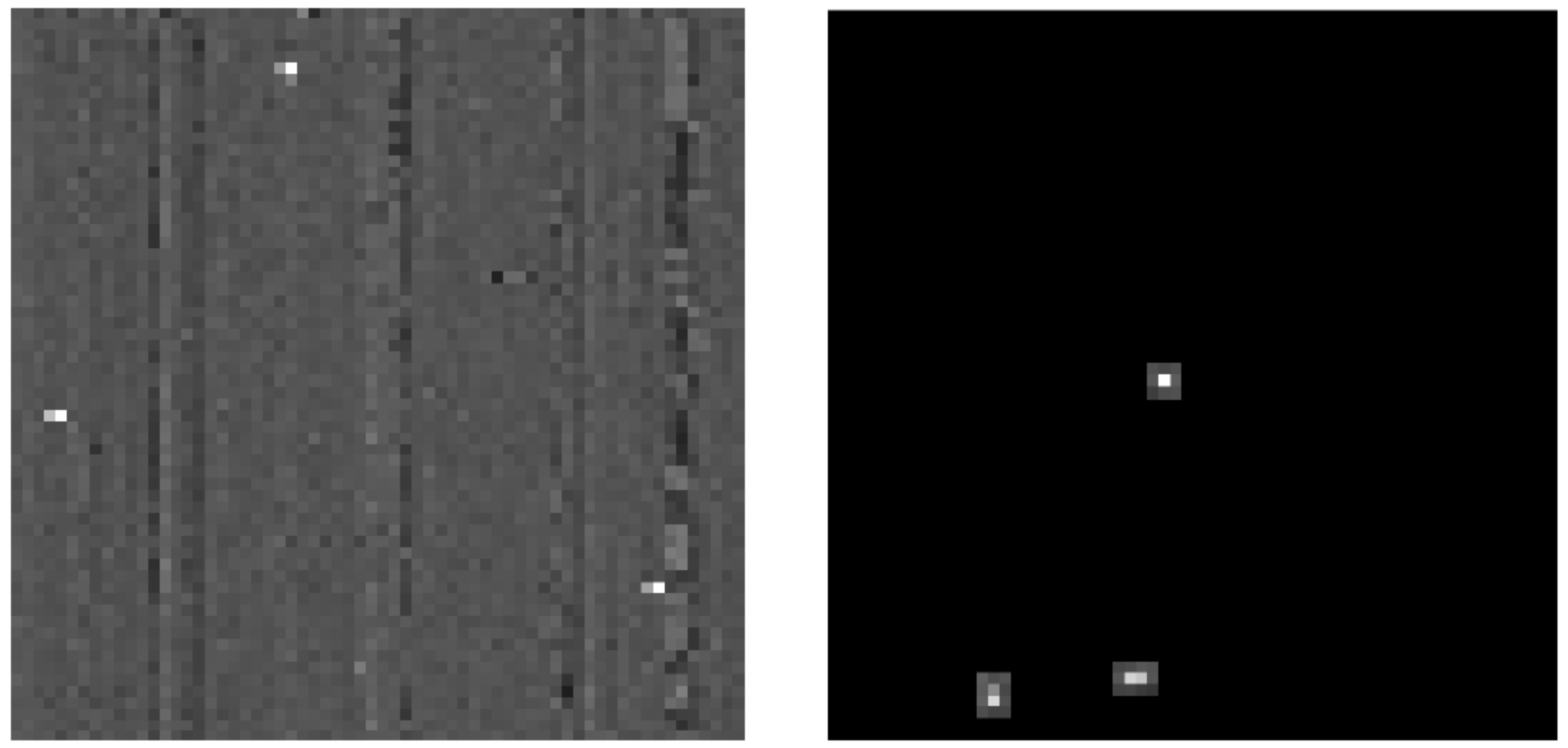}  
     \\
     (a) \hspace{6.6cm} (b)
   \end{tabular}
   \end{center}
   \caption 
   { \label{fig:full_frame_sparse} 
Example images of Full Frame Readout mode (a) where the comparator is set below the noise floor and all 64x64 pixels are read out, and Sparse 3x3 Readout mode (b) where the comparator is set above the read noise floor and only pixels that contain X-ray signal along with the 3x3 region around them are read out.  An $^{55}$Fe X-ray source was used to expose the detector to X-rays.  The central event in the Sparse 3x3 Readout mode image is a single pixel event with X-ray signal in the central pixel and no signal detected in the surrounding pixels (which have levels consistent with noise), while the bottom two events have X-ray signal present in two adjacent pixels, thus leading to a read out of two overlapping 3x3 regions.  The vertical artifacts in the Full Frame Readout image are fixed pattern noise due to the engineering grade of these detectors.  } 
   \end{figure}

 Full Frame Readout mode is a special case of Sparse Readout, where the comparator is set below the noise floor and the entire 64x64 frame is read out.  Example images of Full Frame Readout mode and Sparse 3x3 Readout mode can be seen in Figure \ref{fig:full_frame_sparse}. We will refer to the two modes; Full Frame Readout mode and Sparse 3x3 Readout mode, throughout this paper.


The Speedster-EXD detector can operate at a frame rate up to 10 kHz and has the ability to turn off bad pixels so they are not read out in Full Frame or Sparse Readout mode.  Each Speedster detector has an optical blocking aluminum filter deposited directly on the silicon absorbing layer. The detector is designed to support a count rate of 100,000 counts/s. For a larger 1024 x 1024 pixel$^{2}$ detector, this count rate can be retained by adding multiple channels to the circuitry and readout.  A 550 x 550 pixel$^{2}$ version of the detector is currently being fabricated that retains a count rate capability of up to 100,000 counts per second.
\section{Experiment Setup}
\label{exp_set}
Two test stands were used to cool the Speedster detectors and expose them to X-rays.   Both test stands are described below.

\begin{figure}
   \begin{center}
   \begin{tabular}{c}
   \includegraphics[height=8.5cm]{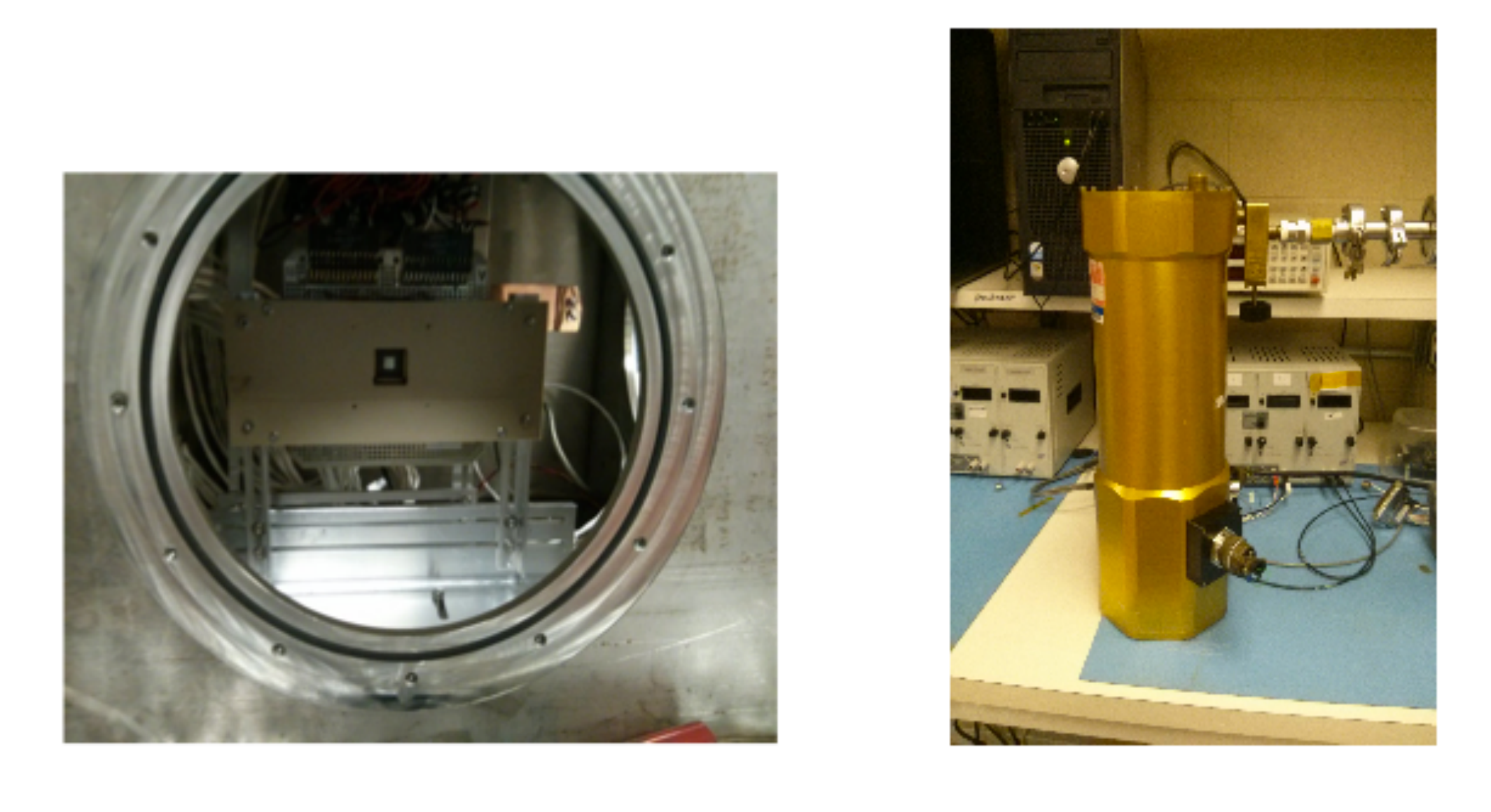}
    \\
    \hspace{1.5cm} (a) \hspace{7.3cm} (b)
   \end{tabular}
   \end{center}
   \caption 
   { \label{fig:speedster_in_cube} 
(a) Speedster-EXD detector mounted inside the cube chamber. The plastic faceplate shown is 15.3 cm x 6.4 cm. (b) IR Labs HDL-5 dewar used for Speedster characterization. The IR Labs dewar is 50.8 cm tall and has a diameter of 17.65 cm.    } 
   \end{figure}

The cube test stand was used to characterize HAWAII HCDs, and it was modified for the Speedster-EXD detector package.  Figure \ref{fig:speedster_in_cube}a shows a Speedster detector mounted inside the cube chamber.  

The cube chamber was evacuated to a pressure of 10$^{-6}$ Torr, allowing the detector to be safely cooled to temperatures as low as 150 K, using liquid nitrogen and a copper cold finger. A radioactive 6.8 millicurie $^{55}$Fe source was used as an X-ray source; it produces Mn K$\alpha$ and K$\beta$ X-rays at 5.89 keV and 6.4 keV.   This source was placed 10 cm away from the detector and produced  $\sim$3 photons/s/pixel.  Two 500 $\mu$Ci radioactive polonium sources that emits alpha particles were also used, which in turn excite an aluminum disc to produce Al K$\alpha$ X-rays at 1.49 keV. This source was also placed 10 cm away from the detector and produced  $\sim$0.3 photons/s/pixel.

A customized IR Labs HDL-5 dewar was used to characterize gain variation, seen in Figure \ref{fig:speedster_in_cube}b.  Using this dewar,  the detector was kept cold for longer periods of time, which was necessary for the gain variation measurements.  The 6.8 millicurie $^{55}$Fe source was used in this setup and was placed 3 cm away from the detector, producing $\sim$30 photons/s/pixel.  As will be discussed in Section \ref{gain_var}, we needed to keep the detectors cold for longer periods of time to capture the millions of images needed to achieve good statistics for the gain variation measurement.  
The HDL-5 dewar was evacuated to a pressure of 10$^{-6}$ Torr.  The detector, within the dewar, was cooled to 150 K using liquid nitrogen, and the $^{55}$Fe source was used for testing. 

Using each of these test stands, two Speedster-EXD detectors were tested and are referred to by their serial numbers; FPA17014 and FPA17017.  Both detectors have an aluminum-coated 100 $\mu$m thick silicon absorbing layer bump bonded to identical ROICs with all of the features described in Section \ref{sect:speedster}. The aluminum optical blocking filter on both detectors is 500 ${\rm \AA}$ thick.  The detectors are fully depleted and run at a bias voltage of 15 V.  The Speedster-EXD detectors are controlled using a camera interface board (Detector Interface Block, DIB) developed by TIS, a Matrox frame grabber, and specially designed software.  The software is designed for the user to control the mode of the Speedster detector (i.e. Sparse Single Pixel Readout Mode, Sparse 3x3 Readout Mode, or Full Frame Readout Mode), the comparator threshold, the exposure time, and other timing parameters.     

\section{Data Reduction}
\label{data_red}
The cube test stand, described in Section \ref{exp_set}, was used to obtain read noise, dark current, IPC, and energy resolution measurements.   Data were taken at a 1 kHz frame rate and in two different modes; Full Frame Readout Mode and Sparse 3x3 Readout Mode.   As mentioned in Section \ref{sect:speedster}, Full Frame Readout mode has the comparator set below the noise floor so all 64x64 pixels are read out (Figure \ref{fig:full_frame_sparse}a), and Sparse 3x3 Readout mode has the comparator set above the noise floor so only pixels that contain charge are read out along with the 3x3 region around them (Figure \ref{fig:full_frame_sparse}b).  Each mode required its own data reduction technique.

In \textit{Full Frame Readout mode}, the pixels that did not contain signal above the noise level in each image were used and averaged together to construct a bias image.  The constructed bias image was then subtracted from each X-ray image.   Swift XRT grade definitions from Photon Counting Mode\cite{2005SSRv..120..165B} were used to properly characterize the events found in the bias subtracted images.  These grading schemes will be discussed further in Section \ref{sect:evt_grading}.  

In \textit{Sparse 3x3 Readout mode}, the pixels that do not contain charge are not read out, so they could not be used to construct a bias frame.  Instead, the comparator was moved below the noise floor (Full Frame Readout Mode), and 1000 dark frames were taken with no X-ray source and averaged together to create a bias frame.  This bias frame was then subtracted from the Sparse 3x3 Readout images.   The same Swift XRT grade definitions were used to grade and characterize the X-ray events.

\section{Analysis and Results}
Once the data were bias subtracted in each mode, each parameter outlined in Section \ref{chara_astronomy} was characterized on both Speedster-EXD detectors.  

\subsection{Interpixel Capacitance Crosstalk}
\label{sect:ipc}
As discussed in Section \ref{ipc_exp}, IPC has been a significant problem for HAWAII HCDs, and special event recognition algorithms had to be used to properly characterize X-ray events on those detectors \cite{2013NIMPA.717...83P, 2010SPIE.7742E..0RB}.  In order to enable the use of standard CCD grading schemes for the Speedster-EXD detectors, we must first characterize the IPC on the Speedster-EXD and determine if it is negligibly small.  

IPC characterization data were taken using the $^{55}$Fe source in Full Frame Readout mode at 150 K, and the images were bias subtracted as explained in Section \ref{data_red}.   X-ray events were found using the criteria that the pixel value exceeded 5$\sigma$ (where $\sigma$ is the measured read noise) and was also a local maximum.  The paired-pixel method described in Prieskorn et. al 2013\cite{2013NIMPA.717...83P} was implemented to measure IPC. This method measures the standard deviation of the pixels above and below the central pixel and pixels on either side of the central pixel of the X-ray event.  If the standard deviation of the above and below pixels and the standard deviation of the side pixels are less than the read noise, the event is classified as suitable for measuring IPC, since the symmetry of the IPC event shows that any signal in the surrounding pixels must not arise from X-ray induced charge in two pixels; i.e. any signal in the surrounding pixels must come from IPC.  The goal of this method is to find single pixel events that are solely affected by IPC.  This method eliminates ``split'' events where charge spreading occurred due to the X-ray being detected near the edge of a pixel.  The single pixel events were averaged and normalized to calculate the 3x3 kernel seen in Table \ref{tab:ipc}.





\begin{table}[h]

\begin{center}
\begin{tabular}{|c|c|c|}

\multicolumn{3}{c}{{\bf FPA17014 IPC Measurement}}\\
\hline
\rule[-1ex]{0pt}{3.5ex} 0.002 $\pm$ 0.001 &0.003 $\pm$ 0.001& 0.001 $\pm$ 0.001 \\
\hline
\rule[-1ex]{0pt}{3.5ex} 0.003 $\pm$ 0.001&0.983 $\pm$ 0.030&0.003 $\pm$ 0.001\\
\hline
\rule[-1ex]{0pt}{3.5ex} 0.002 $\pm$ 0.001&0.001 $\pm$ 0.001& 0.002 $\pm$ 0.001\\
\hline

\end{tabular}
\vspace{.2cm}
\caption{Charge distribution of single-pixel X-ray events used for IPC measurement for FPA17014.  Single pixel events were found using the paired-pixel method. The events were averaged and normalized to calculate the kernel seen.  Nearly all of the charge is retained in the center pixel, and the IPC is measured to be $<$0.65 \% (95 \% confidence level) in the four adjoining pixels.   }
\label{tab:ipc}
\end{center}
\end{table}
In the Speedster detectors, nearly all of the signal is retained in the center pixel, and the average IPC of the four adjacent pixels is 0.25 $\pm$ 0.2 \%.  We calculate an upper limit of 0.65 \% IPC in one of the four adjoining pixels to a 95 \% confidence level.  This result is a major improvement over the HAWAII HCDs with 18 $\mu$m pitch, which showed 4-9 \% average IPC in the four adjacent pixels, and an improvement over the 36 $\mu$m pitch HCD, which showed 1.7-1.8\% average IPC in the four adjacent pixels\cite{2013NIMPA.717...83P, 2012SPIE.8453E..0FG}.  

\subsection{Event Grading}
\label{sect:evt_grading}
Since the Speedster-EXD detector has been shown to have little to no IPC (see Section \ref{sect:ipc}), X-ray events can be properly characterized using standard CCD grading schemes.  The Swift XRT Photon Counting Mode grading scheme\cite{2005SSRv..120..165B} is used throughout this paper.  After the images were bias-subtracted,  primary and secondary thresholds were implemented to find and grade the events.  The event must be measured above the primary threshold of 5$\sigma$ (where $\sigma$ is the read noise) and must be a local maximum.   A secondary threshold was then implemented between 1$\sigma$ and 3$\sigma$ depending on the X-ray energy (see Section \ref{energy_res}), and if any of the surrounding eight pixels were measured above this threshold, their charge was included in the total event charge.  The events were then graded depending on how many pixels were included in the event.  In this paper,  we use Grade 0 (single pixel) events, where all of the charge is contained in the central pixel, and Grade 1-4 events (singly split events), where event charge is diffused into one of the four adjoining pixels, and therefore, two pixels are used in the event. 

\subsection{Gain}
\label{gain}
Mn K$\alpha$ X-rays were used to calculate the gain for each detector in each gain mode.  Raw spectra were used (i.e. not event graded) in each gain mode at 150 K.  All pixels were used in this calculation and each pixel provided similar number of X-ray counts ($\sim$ 20 photons/pixel).  A Gaussian was fit to the Mn K$\alpha$ peak in each spectrum and the mean in DN was found in each gain mode on each detector.  The e$^{-}$/DN factor was found using the standard conversion factor (3.65 eV/e$^{-}$) \cite{2001sccd.book.....J}.   The calculated gains for each mode on each detector are shown in Tables \ref{tab:fap14_read_noise} and \ref{tab:fap17_read_noise}. 

\subsection{Read Noise}
\label{read_noise}
In order to measure the read noise, 1000 exposures with no X-ray source were taken, using both Speedster-EXD detectors.   The standard deviation of the measured signal in each pixel was calculated, and the measured standard deviations were histogrammed.  The mean of the histogram provided a measurement of the read noise in Digital Number (DN).  The read noise was converted to electrons by using the e$^{-}$/DN factors described in Section \ref{gain}. The resulting read noise measurements for FPA17014 and FPA17017 for all four gain modes are shown in Tables \ref{tab:fap14_read_noise} and \ref{tab:fap17_read_noise}. 

\begin{table}[ht]
\begin{center}
\begin{tabular}{c|c}
\multicolumn{2}{c}{{\bf FPA17014}}\\
\hline
Gain Mode & Read Noise \\
\hline
Gain 0 (0.60 e$^{-}$/DN) & 15.2 $\pm$ 0.4 e$^{-}$ \\
Gain 1 (0.72 e$^{-}$/DN) & 14.2 $\pm$ 0.5 e$^{-}$\\
Gain 2 (1.26 e$^{-}$/DN) &13.0 $\pm$ 0.7 e$^{-}$ \\
Gain 3 (2.63 e$^{-}$/DN) & 16.3 $\pm$ 1.4 e$^{-}$\\

\end{tabular}
\vspace{.2cm}
\caption{Measured read noise for FPA17014 in all four gain modes.}
\label{tab:fap14_read_noise}
\end{center}
\end{table}

\begin{table}[ht]
\begin{center}
\begin{tabular}{c|c}
\multicolumn{2}{c}{{\bf FPA17017}}\\
\hline
Gain Mode & Read Noise \\
\hline
Gain 0 (0.67 e$^{-}$/DN) & 12.3 $\pm$ 0.4 e$^{-}$\\
Gain 1 (0.81 e$^{-}$/DN) & 11.2 $\pm$ 0.5 e$^{-}$\\
Gain 2 (1.44 e$^{-}$/DN) & 13.1 $\pm$ 0.7 e$^{-}$ \\
Gain 3 (2.91 e$^{-}$/DN) & 16.9 $\pm$ 1.4 e$^{-}$\\
\end{tabular}
\vspace{.2cm}
\caption{Measured read noise for FPA17017 in all four gain modes.}
\label{tab:fap17_read_noise}
\end{center}
\end{table}

\subsection{Energy Resolution}
\label{energy_res}
Using the  $^{55}$Fe source, both detectors were characterized in Full Frame Readout mode and in Sparse 3x3 Readout mode using all four gain modes.  Using an $\alpha$-emitting Po source with an Al target, the energy resolution of FPA17014 was also measured at 1.49 keV.  This latter measurement was only done on FPA17014 in Full Frame Readout mode at high gain, due to the long times required to obtain reasonable statistics from the weaker Po source.  The number of counts detected from each source were consistent with the photon rates discussed in Section \ref{exp_set}.  



In Full Frame Readout Mode, using the $^{55}$Fe source, 5$\sigma$ was set as the primary threshold and 3$\sigma$ as the secondary threshold, where $\sigma$ is the measured read noise.  Grade 0 events (single pixel events) and Grade 1-4 events (singly split events) were used in the spectra.  The events were histogrammed and, depending on the detector, different approaches were used to measure the energy resolution.  

\begin{figure}[ht!]
\vspace{1cm}
\center
\includegraphics[width=12cm,height=8cm,scale=.75]{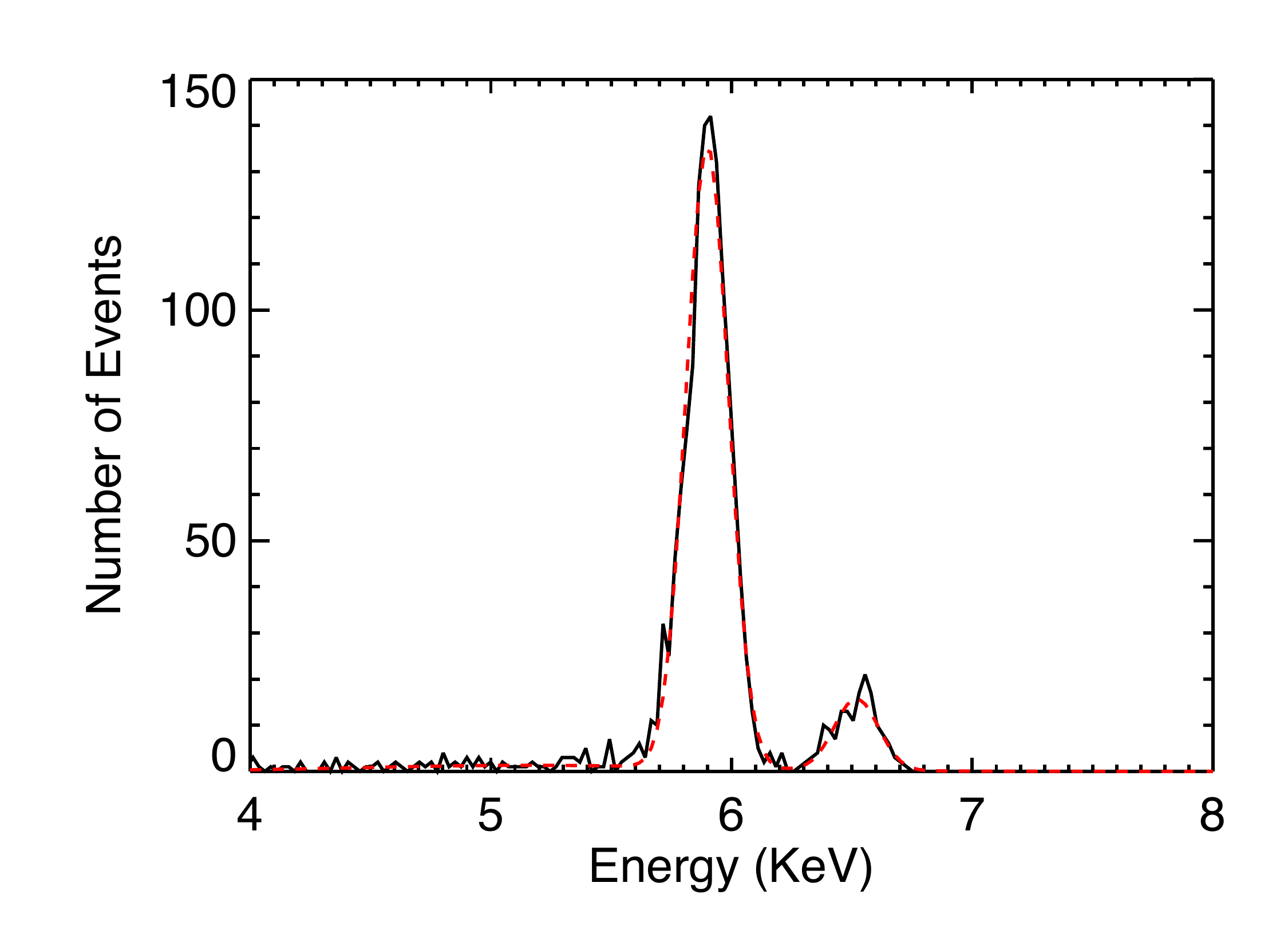}
\caption{FPA17014 Manganese K$\alpha$ and K$\beta$ Grade 0 event spectrum using gain 1 in Full Frame Readout mode.  The measured $\Delta$E/E =  3.5 $\pm$ 0.1 \% with a FWHM = 206 $\pm$ 6 eV.  The two Gaussian fit is shown as a dashed red line. }
\label{gain1_spec}
\end{figure}

 For FPA17014, two Gaussians were fit to each spectrum (one for the Mn K$\alpha$ peak and one for the Mn K$\beta$ peak) and the Full Width Half Max (FWHM) = 2$\sqrt{\rm2ln2}\sigma$ was calculated from the fit of the Mn K$\alpha$ line, and divided by the energy (5.89 keV) to find the $\Delta$E/E percent values.  The FPA17014 spectra were fit well enough by the two Gaussians to derive the FWHM solely from the Gaussian fit parameters (see Figure \ref{gain1_spec}).   The measured FWHM in eV and $\Delta$E/E for FPA17014 in all four gain modes (Gain 0 being the highest) can be found in Table \ref{energy_res_full_14}. The errors were found using the 1$\sigma$ errors on the parameters of the Gaussian fits.

\begin{figure}[htb]
    \centering
    \includegraphics[width=12cm,height=8cm,scale=.75]{{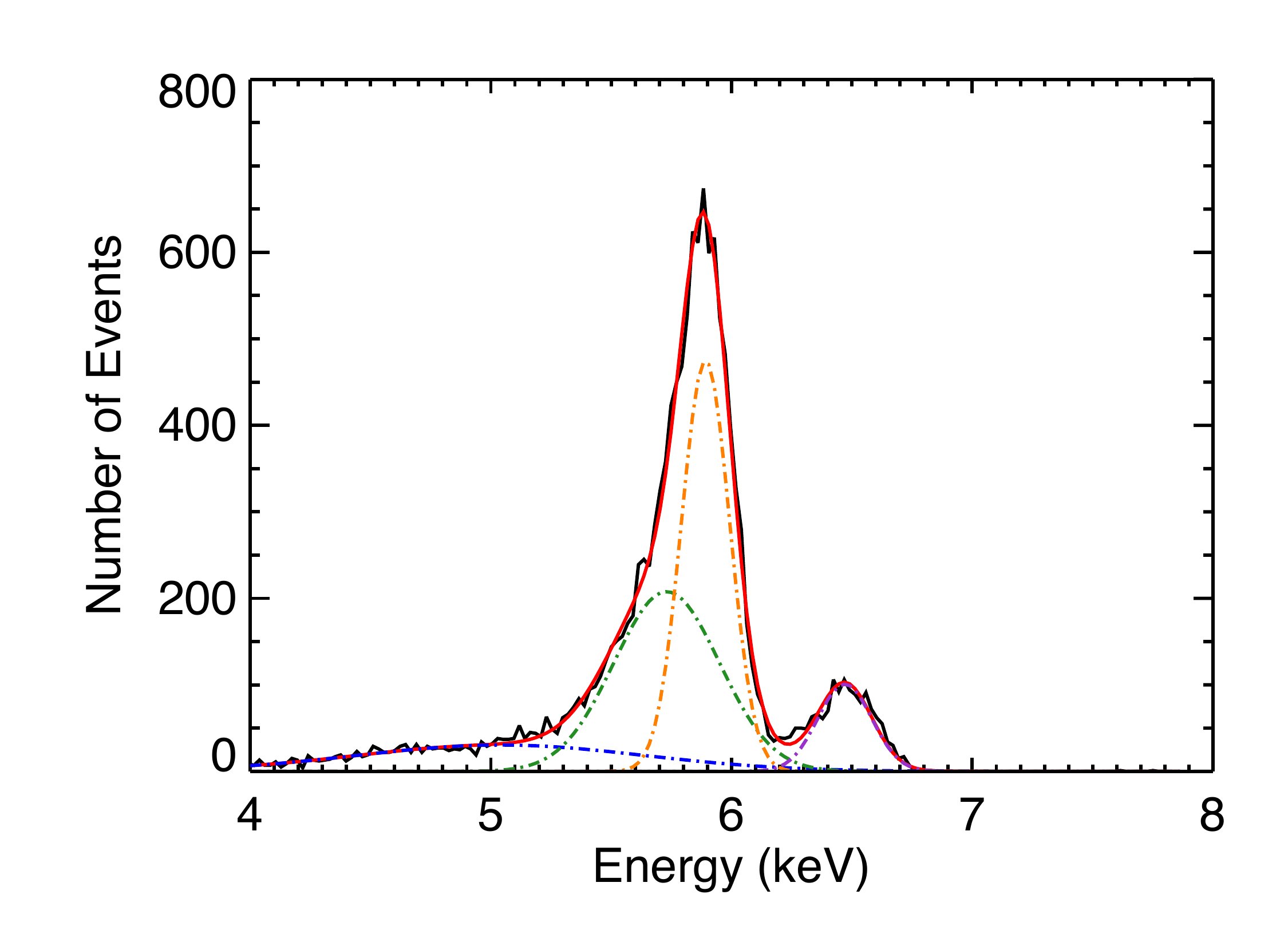}}
    \caption{FPA17017 Manganese K$\alpha$ and K$\beta$ Grade 0 event spectrum using gain 0 in Full Frame Readout mode.  Measured energy resolution for Mn K$\alpha$ $\Delta$E/E = 4.0 $\pm$ 0.2 \% (FWHM = 236 $\pm$ 9 eV) after fitting four Gaussians to the spectrum (dot-dashed lines) and $\Delta$E/E = 5.2 $\pm$ 0.2 \% (FWHM = 304 $\pm$ 10 eV) after measuring the width at half max. This four Gaussian fit was needed to properly fit the measured spectrum ($\chi^{2}_{Reduced}$ $<$ 1). }
    \label{Ch6-figure:FPA17017_gain0_grade0_spec}
\end{figure}

The spectra from FPA17017 were skewed, which we attributed to gain variation on the detector.  This result led us to carry out gain variation measurements on both detectors (see Section \ref{gain_var}).  Due to the large gain variation seen on FPA17017, two approaches were used to find the FWHM due to the large gain variation seen in the detector.  The FWHM was measured by fitting four Gaussians to the spectrum, as shown in Figure \ref{Ch6-figure:FPA17017_gain0_grade0_spec}, and using the measured FWHM of the Gaussian with mean equal to the Mn K$\alpha$ peak of the spectrum (dot-dashed Gaussian with mean at $\sim$5.89 keV in Figure \ref{Ch6-figure:FPA17017_gain0_grade0_spec}).  The four Gaussian fit accounts for the high gain variation, which adds low gain Mn K$\alpha$ events to the spectrum.  This four Gaussian fit was needed to properly fit the measured spectrum ($\chi^{2}_{Reduced}$ $<$ 1), and we do not attempt to interpret the physical meaning of the four Gaussians.  The purpose of the four Gaussian was to provide a phenomenological fit to properly calculate the width of the spectrum around the line we were intending to measure.   As will be further explained in Section \ref{gain_var}, FPA17017 has many low gain pixels that skew the spectrum.  This measurement represents the energy resolution of the Mn K$\alpha$ line without the events from the lower gain pixels. 

The FWHM was also found by simply measuring the width of the large peak in the spectrum at half max, without making any attempt to remove the contribution from the low gain Mn K$\alpha$ X-rays that contaminate the Mn K$\alpha$ peak due to the gain variations combined with fixed thresholds (see Section \ref{gain_var}).    Table  \ref{energy_res_full_17} shows the measured FWHM in eV and $\Delta$E/E for FPA17017 in all four gain modes using both the four Gaussian fit and spectrum width measurement methods. The errors were found using the 1$\sigma$ errors on the parameters of the Gaussian fits. 

\begin{table}[ht]
\begin{center}
\begin{tabular}{|c|c|c|}
\multicolumn{3}{c}{{\bf FPA17014 (Full Frame Readout)}}\\

\hline
Gain Mode &FWHM (eV) & $\Delta$E/E Gaussian Fit\\
 & Gaussian Fit &Gaussian Fit \\

\hline
Gain 0 (Grade 0 Events)  & 215 $\pm$ 5& 3.7 $\pm$ 0.1 \%\\
Gain 1 (Grade 0 Events)  & 206 $\pm$ 6& 3.5 $\pm$ 0.1 \%\\
Gain 2 (Grade 0 Events) & 242 $\pm$ 5& 4.1 $\pm$ 0.1 \%\\
Gain 3 (Grade 0 Events)  & 283 $\pm$ 7 & 4.8 $\pm$ 0.1 \%\\
\hline
Gain 0 (Grade 0-4 Events)  & 292 $\pm$ 5 & 5.0 $\pm$ 0.1 \%\\
Gain 1 (Grade 0-4 Events)  & 276 $\pm$ 6 & 4.7 $\pm$ 0.1 \%\\
Gain 2 (Grade 0-4 Events) & 292 $\pm$ 5 & 5.0 $\pm$ 0.1 \% \\
Gain 3 (Grade 0-4 Events) & 323 $\pm$ 8 & 5.5 $\pm$ 0.1 \% \\
\hline

 \end{tabular}
  \vspace{.1cm}
 \caption{Measured Mn K$\alpha$ (5.89 keV) energy resolution for the FPA17014 Speedster-EXD detector in all four gain modes (Gain 0-3) in Full Frame Readout mode using Grade 0 events (single pixel events) and Grade 0-4 events (single pixel events and singly split events). The measurements were made by fitting a Gaussian to the Mn K$\alpha$ peak.  }
\label{energy_res_full_14}
\end{center}
\end{table}

\begin{table}[ht]
\begin{center}
\begin{tabular}{|c|c|c|c|c|}
\multicolumn{5}{c}{{\bf FPA17017 (Full Frame Readout)}}\\
\hline
Gain Mode &FWHM (eV) & $\Delta$E/E & FWHM (eV) & $\Delta$E/E\\
 & Spectrum Width& Spectrum Width & Gaussian Fit & Gaussian Fit\\
\hline
Gain 0 (Grade 0 Events) & 304  $\pm$ 10 & 5.2  $\pm$ 0.2 \%& 236 $\pm$ 9 & 4.0 $\pm$ 0.2 \%\\
Gain 1 (Grade 0 Events) & 310  $\pm$ 5 & 5.3  $\pm$ 0.1 \%& 236 $\pm$ 6 & 4.0 $\pm$ 0.1 \%\\
Gain 2 (Grade 0 Events)  & 331  $\pm$ 10 & 5.6  $\pm$ 0.2 \%& 236 $\pm$ 8 & 4.0 $\pm$ 0.1 \%\\
Gain 3 (Grade 0 Events) & 312  $\pm$ 20 & 5.3  $\pm$ 0.3 \%& 271 $\pm$ 12 & 4.6 $\pm$ 0.2 \%\\
\hline
Gain 0 (Grade 0-4 Events) & 431 $\pm$ 5& 7.3  $\pm$ 0.1 \% & 330 $\pm$ 12 & 5.6 $\pm$ 0.2 \%\\
Gain 1 (Grade 0-4 Events) & 423 $\pm$ 5& 7.2  $\pm$ 0.1 \%& 318 $\pm$ 9 & 5.4 $\pm$ 0.2 \%\\
Gain 2 (Grade 0-4 Events) & 453 $\pm$ 10& 7.7  $\pm$ 0.2 \%& 324 $\pm$ 12 & 5.5 $\pm$ 0.2 \%\\
Gain 3 (Grade 0-4 Events) & 389 $\pm$ 20& 6.6  $\pm$ 0.3 \%& 336 $\pm$ 14 & 5.7 $\pm$ 0.2 \%\\
\hline
\end{tabular}
 \vspace{.1cm}
\caption{Measured Mn K$\alpha$ (5.89 keV) energy resolution for the FPA17017 Speedster-EXD detector in in all four gain modes (Gain 0-3) in Full Frame Readout mode using Grade 0 events (single pixel events) and Grade 0-4 events (single pixel events and singly split events). Spectrum Width refers to the FWHM measurement made by simply measuring the spectrum width at half max of the large Mn K$\alpha$ peak and Gaussian Fit refers to the FWHM measurement made by fitting four Gaussians to the spectrum (see text).}
\label{energy_res_full_17}
\end{center}
\end{table}

The best energy resolution is measured to be $\Delta$E/E = 3.5 $\pm$ 0.1 \% (FWHM = 206 $\pm$ 6 eV) on FPA17014 in Gain 1 mode (second highest gain mode) using Grade 0 events.  The fitted spectrum can be seen in Figure \ref{gain1_spec}.  This energy resolution measurement is the best to date at 5.89 keV on an X-ray HCD. The discrepancy in the energy resolution measurements on FPA17014 and FPA17017, despite the detectors having similar read noise (see Tables \ref{tab:fap14_read_noise} and \ref{tab:fap17_read_noise}), further motivated gain variation measurements on both detectors (see Section \ref{gain_var}).

\begin{figure}
   \begin{center}
   \begin{tabular}{c}
   \includegraphics[height=7cm]{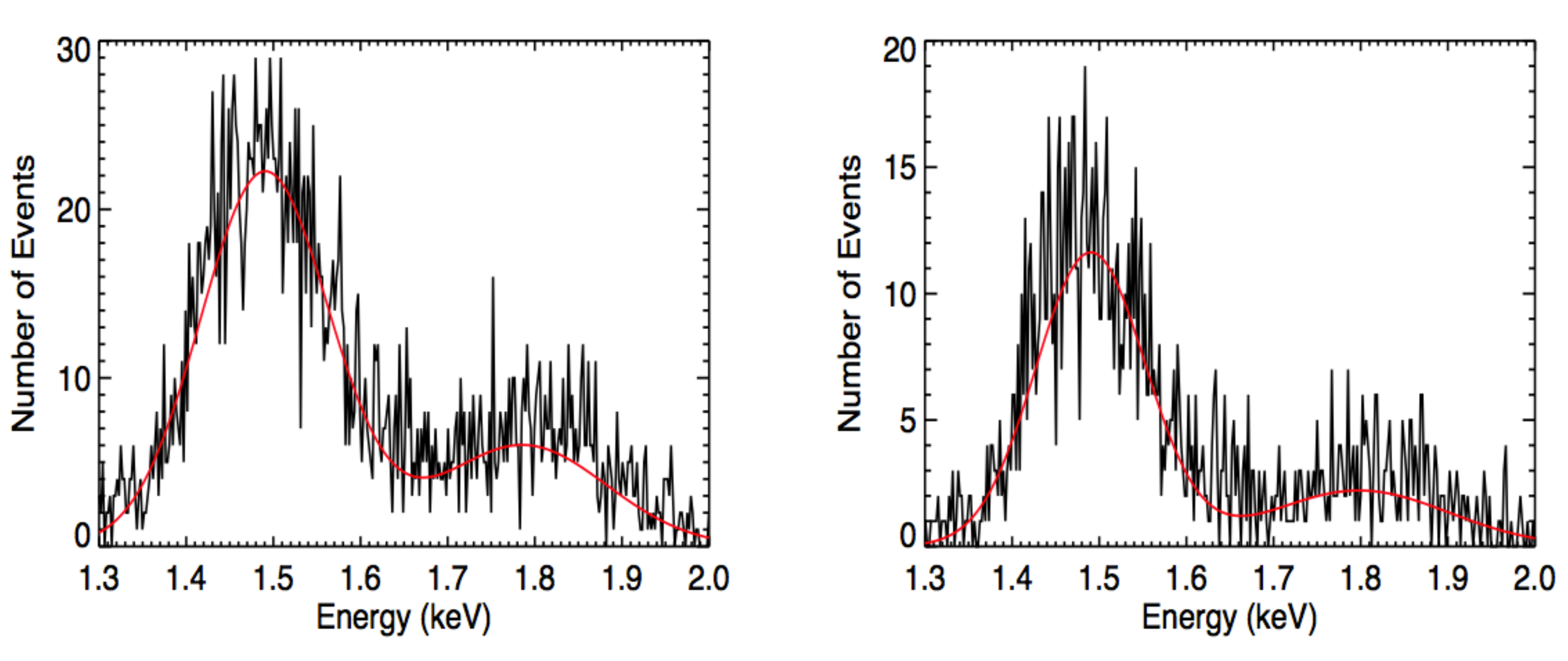}  
     \\
     (a) \hspace{8cm} (b)
   \end{tabular}
   \end{center}
   \caption 
   { \label{fig:al_fits} 
a) Aluminum K$\alpha$ (1.49 keV) spectrum taken in gain 0 in Full Frame Readout mode using Grade 0 events on FPA17014 with the secondary threshold set to 2$\sigma$.  The measured $\Delta$E/E = 11.6 $\pm$ 0.3 \% with a FWHM = 172 $\pm$ 5 eV.  The peak near 1.8 keV is from silicon contamination in the aluminum source producing silicon lines. b)  Aluminum K$\alpha$ (1.49 keV) spectrum taken in gain 0 in Full Frame Readout mode using Grade 0 events on FPA17014 with the secondary threshold set to 1$\sigma$. The measured $\Delta$E/E = 10.0 $\pm$ 0.3 \% with a FWHM = 148 $\pm$ 5 eV.  Two Gaussian fit is shown as a red solid line in both spectra. } 
   \end{figure}

A fluorescent aluminum source was used to produce 1.49 keV X-rays.  Due to the low activity of the source, and therefore, the exceptional time required for sufficient counts,  only FPA17014 was characterized in Full Frame Readout mode using Gain 0.   The primary threshold was 5$\sigma$, and the secondary threshold was set to 2$\sigma$.   The secondary threshold was lowered on the aluminum source data due to the relationship found between optimal secondary threshold and energy when testing the HAWAII HCDs\cite{2013NIMPA.717...83P}.   The energy resolution was calculated, using the Gaussian fit parameters of the Al K$\alpha$ line, to be $\Delta$E/E = 11.6 $\pm$ 0.3 \% (FWHM = 172 $\pm$ 5 eV) using Grade 0 events.  The fitted spectrum can be seen in Figure \ref{fig:al_fits}a. The line seen just below 1.8 keV is from silicon contamination in the aluminum source leading to a production of silicon lines.   


With the secondary threshold set to $\sim$1$\sigma$, the energy resolution was calculated to be $\Delta$E/E = 10.0 $\pm$ 0.3 \% (FWHM = 148 $\pm$ 5 eV; Figure \ref{fig:al_fits}b).  This single-pixel event spectrum has $\sim$50 \% fewer events than the spectrum with a secondary threshold equal to 2$\sigma$, but retains the best Grade 0 events and thereby gives the best energy resolution. This measurement is the best energy resolution to date at 1.49 keV on an X-ray HCD.


For Sparse Readout mode, the 3x3 version was used to properly characterize and grade the events.  A bias frame was constructed using the procedure described in Section \ref{data_red}.  The same 5$\sigma$ and 3$\sigma$ primary and secondary thresholds were used on the  $^{55}$Fe source data, and Grade 0-4 events were used in the spectra.  The same methods for calculating $\Delta$E/E were used on each detector as described in the previous section.   The measured values are shown in Tables \ref{energy_res_sparse_14} and \ref{energy_res_sparse_17}. The best energy resolution is measured to be $\Delta$E/E = 4.2 $\pm$ 0.1 \% with a FWHM = 247 $\pm$ 5 eV on FPA17014 at Gain 0, using Grade 0 events.

\begin{table}[ht]
\begin{center}
\begin{tabular}{|c|c|c|}
\multicolumn{3}{c}{{\bf FPA17014 (3x3 Sparse Readout)}}\\
\hline
Gain Mode & FWHM (eV) & $\Delta$E/E\\
 & Gaussian Fit &Gaussian Fit \\
\hline
Gain 0 (Grade 0 Events) & 247 $\pm$ 5 & 4.2 $\pm$ 0.1 \%\\
Gain 1 (Grade 0 Events) & 277 $\pm$ 6 & 4.7 $\pm$ 0.1 \%\\
Gain 2 (Grade 0 Events) & 353 $\pm$ 5 & 6.0 $\pm$ 0.1 \%\\
Gain 3 (Grade 0 Events) & 377 $\pm$ 23 & 6.4 $\pm$ 0.4 \%\\
\hline
Gain 0 (Grade 0-4 Events)& 330 $\pm$ 6 & 5.6 $\pm$ 0.1 \%\\
Gain 1 (Grade 0-4 Events) & 353 $\pm$ 5 & 6.0 $\pm$ 0.1 \%\\
Gain 2 (Grade 0-4 Events) & 454 $\pm$ 5 & 7.7 $\pm$ 0.1 \%\\
Gain 3 (Grade 0-4 Events) & 683 $\pm$ 15 & 11.6 $\pm$ 0.3 \%\\
\hline
\end{tabular}
 \vspace{.1cm}
\caption{Measured Mn K$\alpha$ (5.89 keV) energy resolution for the FPA17014 Speedster-EXD detector in in all four gain mode (Gain 0-3) in 3x3 Sparse Readout mode using Grade 0 events (single pixel events) and Grade 0-4 events (single pixel events and singly split events). The measurements were made by fitting a Gaussian to the Mn K$\alpha$ peak. }
\label{energy_res_sparse_14}
\end{center}
\end{table}

\begin{table}[ht]
\begin{center}
\begin{tabular}{|c|c|c|c|c|}
\multicolumn{5}{c}{{\bf FPA17017 (3x3 Sparse Readout)}}\\
\hline
Gain Mode & FWHM (eV) & $\Delta$E/E & FWHM (eV) & $\Delta$E/E\\
 & Spectrum Width& Spectrum Width & Gaussian Fit & Gaussian Fit\\
\hline
Gain 0 (Grade 0 Events) &276 $\pm$ 5 & 4.7 $\pm$ 0.1 \%& 247 $\pm$ 5 & 4.2 $\pm$ 0.1 \%\\
Gain 1 (Grade 0 Events) &303 $\pm$ 5& 5.2 $\pm$ 0.1 \%& 236 $\pm$ 5 & 4.0 $\pm$ 0.1 \%\\
Gain 2 (Grade 0 Events) &417 $\pm$ 10&7.1 $\pm$ 0.2 \%& 377 $\pm$ 8 & 6.4 $\pm$ 0.1 \%\\
Gain 3 (Grade 0 Events) & 470 $\pm$ 20& 8.0 $\pm$ 0.3 \%& 395 $\pm$ 22 & 6.7 $\pm$ 0.4 \%\\
\hline
Gain 0 (Grade 0-4 Events)& 327 $\pm$ 5& 5.6 $\pm$ 0.1 \%& 306 $\pm$ 5 & 5.2 $\pm$ 0.1 \%\\
Gain 1 (Grade 0-4 Events) & 365 $\pm$ 5& 6.2 $\pm$ 0.1 \%& 336 $\pm$ 5 & 5.7 $\pm$ 0.1 \%\\
Gain 2 (Grade 0-4 Events) & 517 $\pm$ 10& 8.8 $\pm$ 0.2 \%& 441 $\pm$ 16 & 7.5 $\pm$ 0.3 \%\\
Gain 3 (Grade 0-4 Events) & 742 $\pm$ 20& 12.6 $\pm$ 0.3 \%& 471 $\pm$ 22 & 8.0 $\pm$ 0.4 \%\\
\hline
\end{tabular}
 \vspace{.1cm}
\caption{Measured Mn K$\alpha$ (5.89 keV) energy resolution for the FPA17017 Speedster-EXD detector in in all four gain mode (Gain 0-3) in 3x3 Sparse Readout mode using Grade 0 events (single pixel events) and Grade 0-4 events (single pixel events and singly split events). Spectrum Width refers to the FWHM measurement made by simply measuring the spectrum width at half max of the large Mn K$\alpha$ peak and Gaussian Fit refers to the FWHM measurement made by fitting four Gaussians to the spectrum (see text).  }
\label{energy_res_sparse_17}
\end{center}
\end{table}

\subsection{Dark Current}
\label{sect:dark_current}

The Speedster-EXD detectors were tested at 10 K increments from 260 K to 150 K.  At each temperature,  multiple dark images were taken at different integration times.  The images' integration times ranged from 1 ms to 1 second.   The mean of each image was found at each integration time at each temperature.   A line was fit to the measured means of each image at each temperature to find the slope in DN/s.  The result was converted to e$^{-}$/s using the same e$^{-}$/DN factor discussed in Section \ref{gain}.  Reported errors are 1$\sigma$ errors on the linear fit parameters.  



 The function used to fit the measured dark current is below \cite{2001sccd.book.....J}:
 \begin{equation}
 {\rm Dark \ Current }{(\rm e}^{-}{\rm/s)} = 2.5\times10^{15} P_{s}D_{FM}T^{1.5}exp\bigg(\frac{-E_{g}}{2kT}\bigg)
 \label{Dfm}
 \end{equation}
  where $P_{s}$ is the pixel size in cm$^{2}$, $T$ is the temperature in Kelvin, $D_{FM}$ is the dark current figure of merit at $T$ =  293 K, $k$ is the Boltzmann's constant, and $E_{g}$ (eV) is the silicon band gap energy as a function of $T$.  $E_{g}$ has the form:
  \begin{equation}
 E_{g} = 1.1557 - \frac{7.021\times10^{-4}T^{2}}{1108 + T}
 \end{equation}
The measured dark current at 150 K and the 293 K fit using Equation \ref{Dfm} for each detector are shown in Table \ref{dark_current_table}.  The lowest dark current measured was (4.6 $\pm$ 0.5) $\times$ 10$^{-3}$  e$^{-}$/s/pixel on FPA17017.

\begin{table}[ht!]
\begin{center}
\begin{tabular}{|c|c|c|}

\multicolumn{3}{c}{}\\
\hline
& {\bf 150 K (e$^{-}$/s/pixel)} & {\bf 293 K Fit (e$^{-}$/s/pixel)}\\
\hline
{\bf FPA17014} &(4.9 $\pm$ 0.6 ) $\times$ 10$^{-3}$ & 5.3 $\times$ 10$^{3}$ \\ 
\hline
{\bf FPA17017} &(4.6 $\pm$ 0.5) $\times$ 10$^{-3}$ & 3.3 $\times$ 10$^{3}$ \\ 
\hline
\end{tabular}
\vspace{.2cm}
\caption{Measured dark current values at 150 K and the 293 K fit for both FPA17014 and FPA17017.} 
\label{dark_current_table}
\end{center}
\end{table}

\begin{figure}  
\center
\includegraphics[width = 15cm,  height = 2.5in]{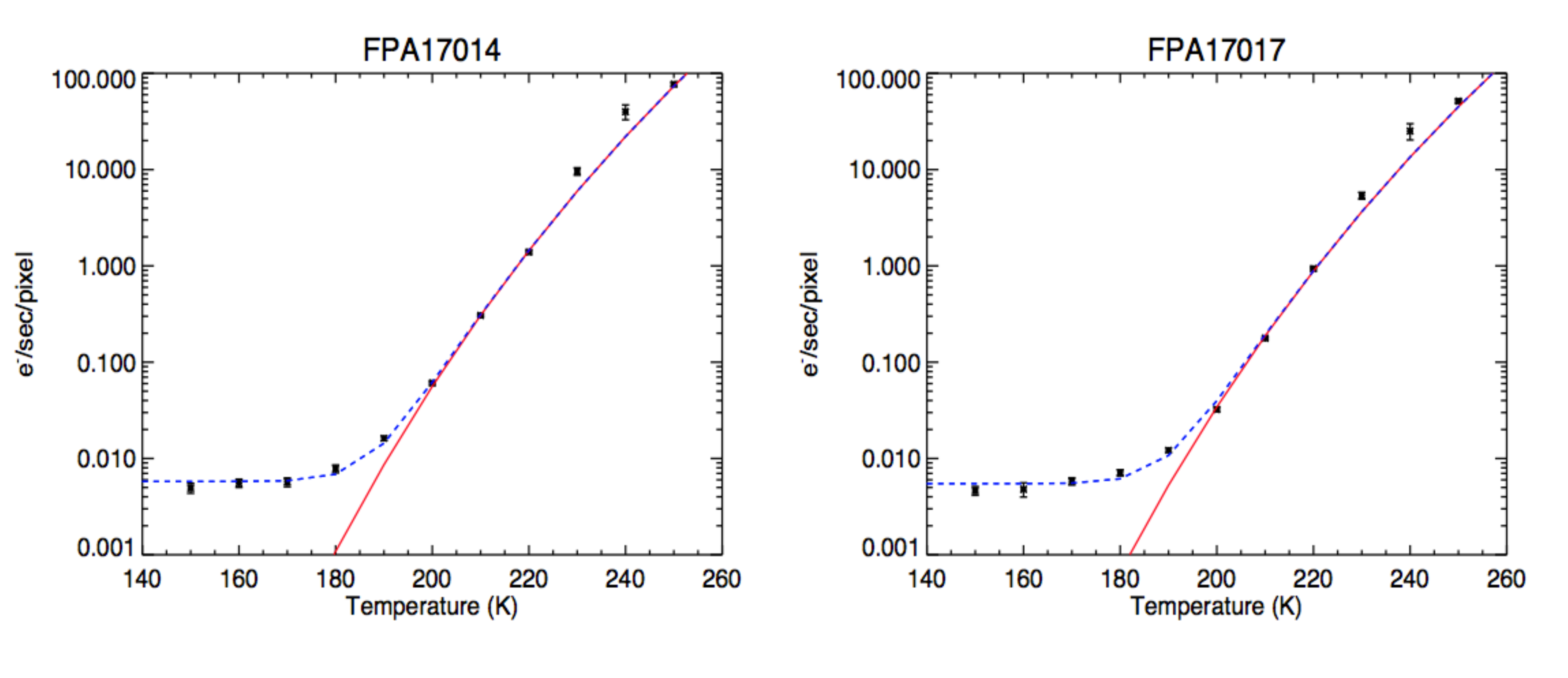}
\caption{FPA17014 and FPA17017 dark current measurements fit with Equation \ref{Dfm} (red solid line) and Equation \ref{Dfm} plus a constant (blue-dash line).  }
\label{dark_current_fit}
\end{figure}

Figure \ref{dark_current_fit} shows the dark current measurements at each temperature and the fit from Equation \ref{Dfm} (solid red line) for both FPA17014 and FPA17017.   Both detectors follow this equation above 200 K, although the data deviate from the curve at 230 K and 240 K.  A constant was added (blue dash line) to fit the dark current measurements below 200 K.  The added constant could be indicative of a separate source of dark current below 200 K (see Prieskorn et. al 2013\cite{2013NIMPA.717...83P}). Both detectors show minimal reduction in dark current below $\sim$170 K.

\subsection{Gain Variation}
\label{gain_var}
\begin{figure}[!t]
    \centering
    \includegraphics[width=12cm,height=8cm,scale=.75]{{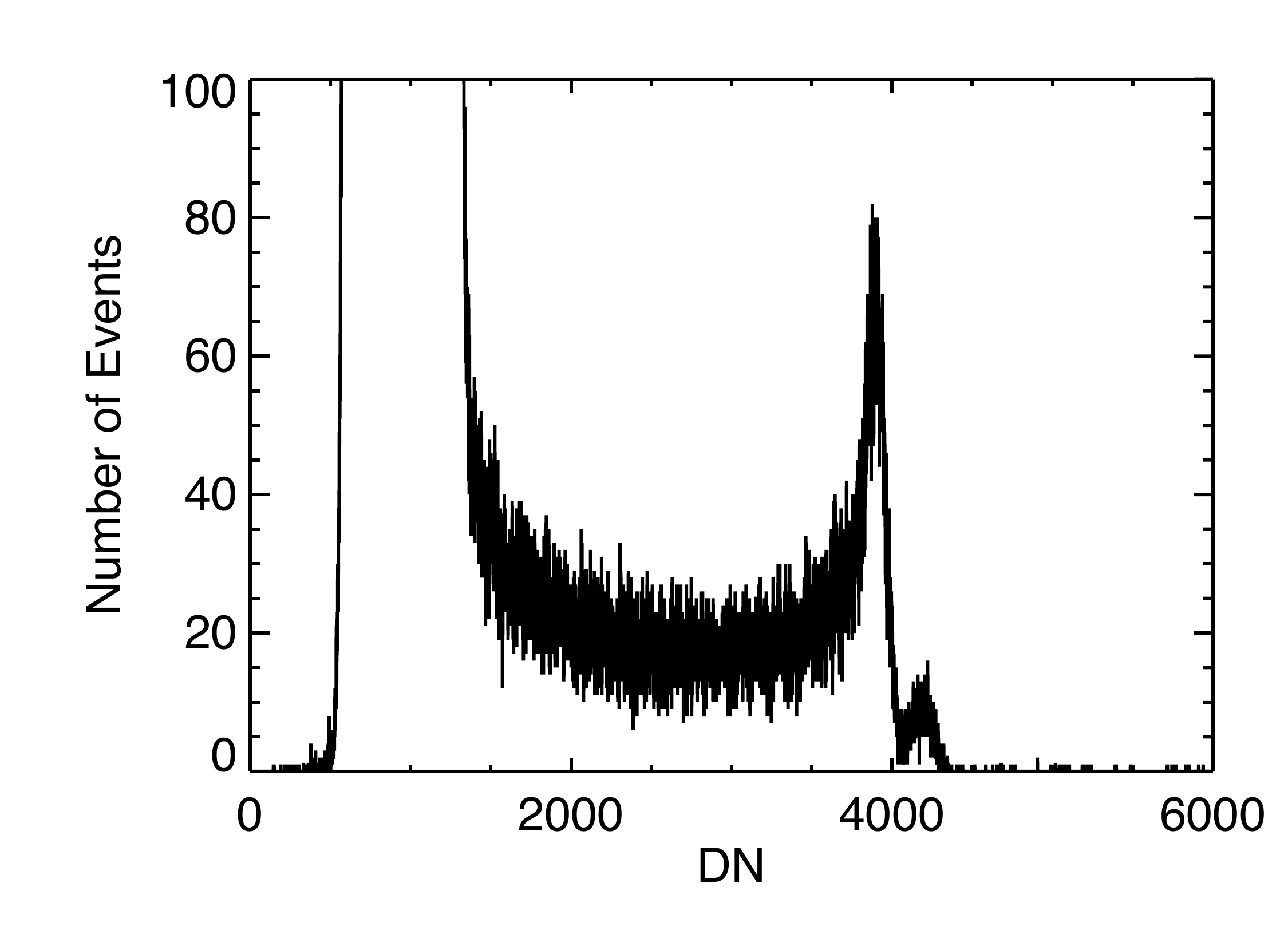}}
    \caption{Raw (unprocessed) spectrum of one pixel on FPA17017 after 30 million images.  The Mn K$\alpha$ peak contains around 10,000 events. The noise peak is located at around $\sim$960 DN and the Mn K$\alpha$ and K$\beta$ peak are located at approximately 3800 DN and 4200 DN. The left (low energy) tail of the Mn K$\alpha$ spectrum is caused by split events that were detected in the pixel. }
    \label{gain_var_pix_example}
\end{figure}

The detectors  were installed in the HDL-5 dewar for the gain variation measurements.   Data were collected until $\sim$10,000 events/pixel were detected.  The gain variation data were taken in Full Frame Readout mode using Gain 0 on both detectors.  The captured images were histogrammed  as they were taken to build up a noise peak and a Mn K$\alpha$ and K$\beta$ peak for each pixel.  An example raw spectrum from one pixel on FPA17017 is shown in Figure \ref{gain_var_pix_example}.   The left (low energy) tail of the Mn K$\alpha$ spectrum in Figure \ref{gain_var_pix_example} is caused by split events (Grade 1-4) that were detected in the pixel.  A $\sim$960 DN offset was added to the data to more easily fit the zero peak in the data reduction process.

The noise peak, Mn K$\alpha$, and the Mn K$\beta$ peaks were all well fit by a Gaussian distribution in each pixel.  A gain measurement for each pixel was found by subtracting the mean of the noise peak from the mean of the Mn K$\alpha$ peak (in DN units).  A gain variation map was then created for each detector. Bad pixels were eliminated from the gain variation measurement because they produce bad data and would be masked out when used in any scientific application.  Bad pixels are defined as those that are dead or those that have drastically high or low gain (e.g. Mn K$\alpha$ peak resides near the full-well or near the noise peak).  The gain variation for each detector was found by calculating the standard deviation of the gain variation map for each detector (with bad pixels removed) and dividing it by the mean.  The errors were calculated by finding the one sigma errors on the mean of the Gaussian fit for each pixel and calculating the mean of the entire map.

\subsubsection{FPA17014}
\begin{figure}[htb]
    \centering
    \includegraphics[width=12cm,height=8cm,scale=.75]{{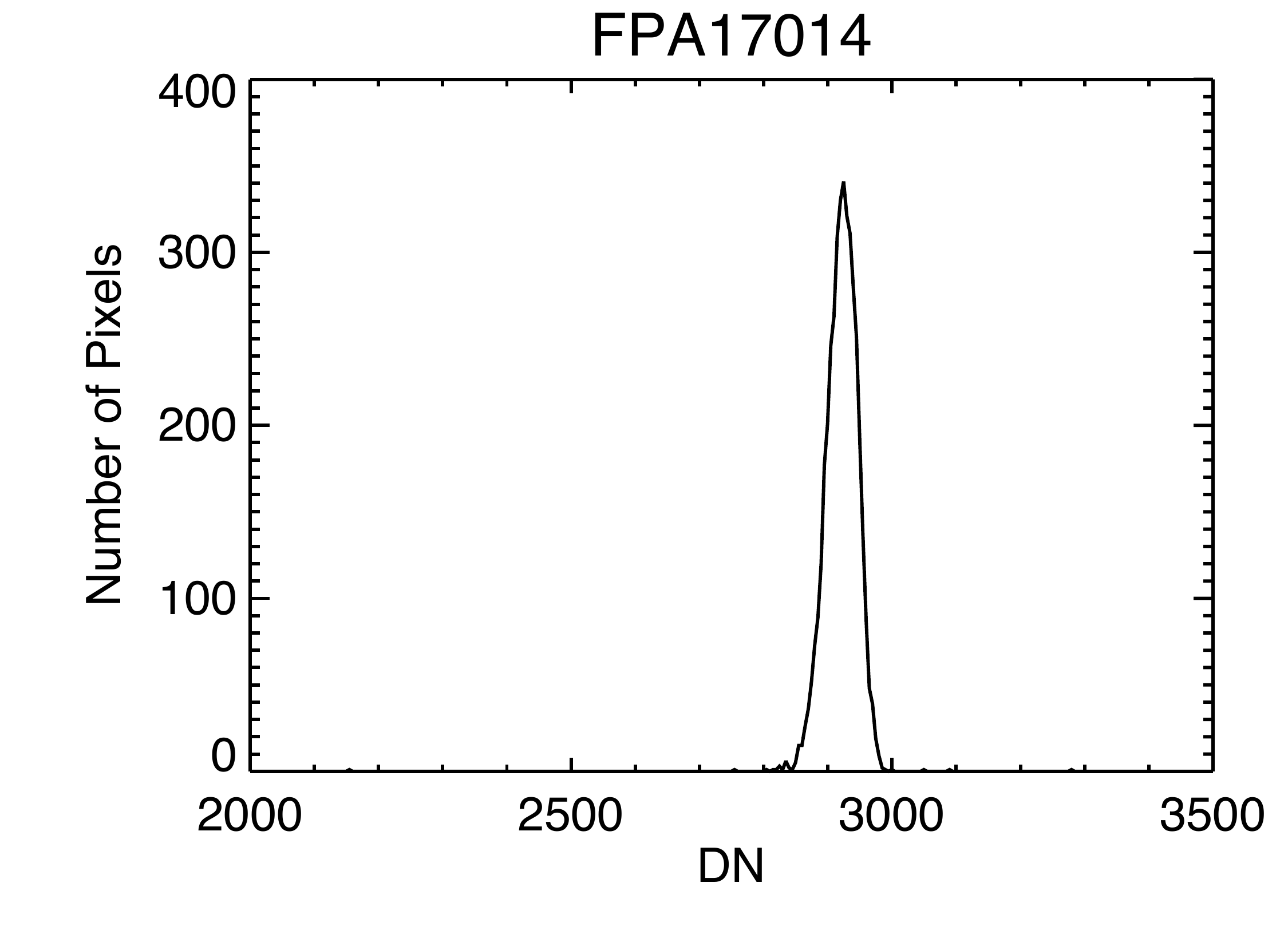}}
    \caption{Histogram of pixels from FPA17014 gain variation map.}
    \label{Ch6-figure:FPA17014_gain_var_hist}
\end{figure}

Approximately $\sim$2 \% of the pixels on the detector were classified as bad and excluded from the gain variation measurement, while all other pixels were included in the calculation. Gain variation was measured to be 0.80 $\pm$ 0.03 \% on FPA17014.  Figure \ref{Ch6-figure:FPA17014_gain_var_hist} shows a histogram of the gain variation map, where the zero-subtracted Mn K$\alpha$ mean of each pixel is histogrammed.      
 
  \begin{figure}[htb]
    \centering
    \includegraphics[width=12cm,height=8cm,scale=.75]{{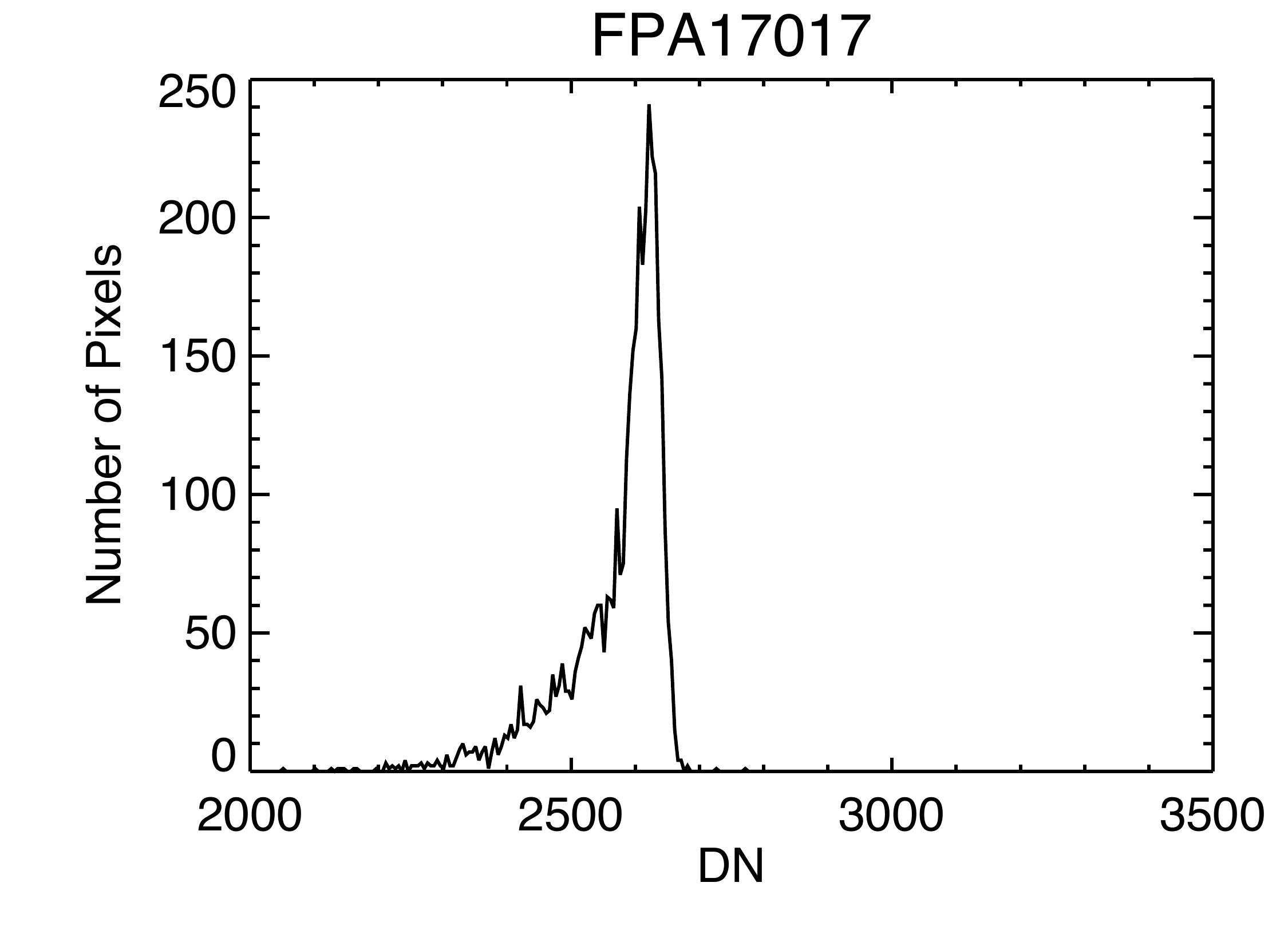}}
    \caption{Histogram of pixels from FPA17017 gain variation map.}
    \label{Ch6-figure:FPA17017_gain_var_hist}
\end{figure}
 
 \subsubsection{FPA17017}
Approximately $\sim$3 \% of the pixels on the detector were classified as bad and excluded from the gain variation measurement, while all other pixels were included in the calculation. Gain variation was measured to be 3.40 $\pm$ 0.03 \% on FPA17017.   Figure \ref{Ch6-figure:FPA17017_gain_var_hist} shows a histogram of the gain variation map, where the zero-subtracted Mn K$\alpha$ mean of each pixel is histogrammed.

\begin{figure}[!t]
    \centering
    \includegraphics[width=8cm,height=8cm,scale=.75]{{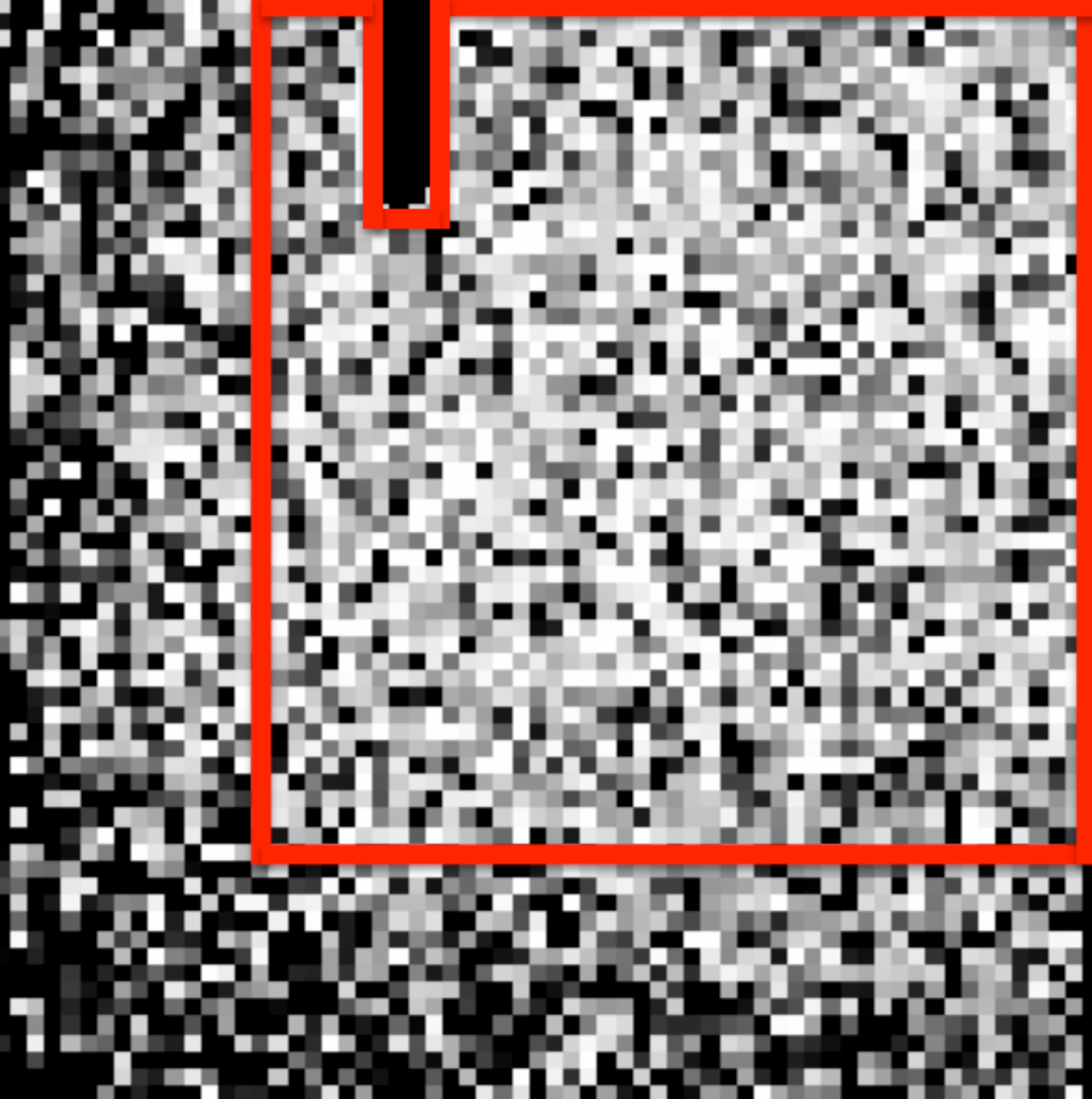}}
    \caption{FPA17017 gain variation map.  Brightness corresponds to the gain measured in the pixel.  Lower gain pixels (dark) are predominately located in the bottom and left side of the detector.  Red box highlights region used to reduce gain variations from 3.4 \% to around 2 \%. The block of black pixels in upper part of the detector is dead pixels, which were not used in the gain variation calculation.  }
    \label{Ch6-figure:FPA17017_gain_var_map}
\end{figure}

 Clearly the histogram of FPA17017 has a much larger low energy tail than that of FPA17014, causing the much larger gain variation.  However, the main peaks in both FPA17014 and FPA17017 have similar widths.  Figure \ref{Ch6-figure:FPA17017_gain_var_map} displays the spatial gain map for FPA17017, where brightness corresponds to higher gain.  A spatial relation emerges where pixels with low gain are located predominantly in the bottom and left side of the detector.  If a boundary of 15 pixels along the left and bottom is excluded (red box in Figure \ref{Ch6-figure:FPA17017_gain_var_map}), the gain variation drops to around 2 \%.   This drop in gain variation shows that a significant portion of the low gain pixels are located in the left side and bottom side of the detector.  The precise reason behind the gain variation discrepancy between the two engineering grade detectors is unknown.

\subsection{Monte Carlo Simulation}
After the gain variation on both Speedster detectors was characterized, a Monte Carlo simulation was run to test whether the magnitude of the gain variation could account for the fact that the measured $\Delta$E/E is larger than that expected from read noise alone (see Section \ref{energy_res}).  

Prior to performing this test, one additional noise factor needs to be taken into consideration.  When taking the gain variation data using the HDL-5 dewar, an increase in noise was identified, relative to the earlier read noise measurement.  A read noise calculation was done on each detector in the HDL-5 dewar, and both detectors showed an increase compared to the measured value reported in Section \ref{read_noise}.   This noise increase was later found to be from the system vacuum pumps, and will be referred to it as ``pump noise''.  For the simulations in this section, we refer to the sum of the pump noise and the read noise as $\sigma_{RN+Pump}$.

 Assuming negligible dark current and gain variation, the theoretical limit of energy resolution on a detector can be found using the equation:
 \begin{equation}
 \frac{\Delta E}{E} (eV) = \frac{2.354 \omega}{E} \sqrt{\frac{FE}{\omega} + (\sigma_{RN+Pump})^{2}} 
 \label{e_res_theo_limit}
 \end{equation}
 where $\omega$ = 3.65 eV/e$^{-}$, $F$ is the Fano factor (0.11 for silicon) \cite{1947PhRv...72...26F, 1997NIMPA.399..354L, 2008NIMPA.584..436M}, $E$ is the energy of the incident X-ray photon, and $\sigma_{RN+Pump}$ is the measured read noise plus vacuum pump noise.  It is worth noting that the standard Fano factor value quoted by authors is 0.11 for silicon\cite{1947PhRv...72...26F, 1997NIMPA.399..354L, 2008NIMPA.584..436M}, while one group of authors report a higher value of 0.15\cite{2002NIMPA.491..437O}.   Slightly different approaches were taken to model the gain variation for each detector. 
 
 
  \subsubsection{FPA17014}
 To model the gain variation on FPA17014, a Gaussian distribution of events was created with the mean equal to the empirically measured Mn K$\alpha$ mean of 1614 e$^{-}$ and the standard deviation equal to the combination, in quadrature, of the deviations from the Fano factor term and the measured $\sigma_{RN+Pump}$.  
 
A 64x64 pixel gain variation map was created starting with a random Gaussian distribution with a mean at zero percent and using the measured gain variation of FPA17014 (0.80 \%) as the standard deviation.  The Gaussian distribution of events was convolved through the gain variation map, and the resulting events were histogrammed to create a modeled spectrum.  The spectrum was then fit with a Gaussian to calculate the modeled energy resolution.  
 
Applying Equation \ref{e_res_theo_limit}, using the measured $\sigma_{RN+Pump}$ in Gain 0 mode (27 e$^{-}$), a theoretical energy resolution limit of $\Delta$E/E = 4.4 \% was calculated for FPA17014.  This theoretical limit is much higher than the measured value presented in Table \ref{energy_res_full_14} due to the increase in noise seen from the vacuum pumps when using the HDL-5 dewar.  After convolving the modeled spectrum, with the above increased noise, through a gain variation map with gain variation equal to 0.80 \%, the modeled spectrum was measured to have a $\Delta$E/E = 4.8 $\pm$ 0.2 \%.  This modeled value is consistent with the measured value with the increased noise, $\sigma_{RN+Pump}$, of $\Delta$E/E = 5.0 $\pm$ 0.2 \%. 

 
Assuming the gain variation was the same when we measured the lower read noise in the cube test stand (see Section \ref{read_noise}), the measured Gain 0 read noise (15.2 e$^{-}$) can be combined with this gain variation to obtain the expected $\Delta$E/E in this set up.  This results in a value of $\Delta$E/E = 3.5 $\pm$ 0.2 \%, which is consistent with the measured value of $\Delta$E/E = 3.7 $\pm$ 0.2 \% as reported in Section \ref{energy_res}.  
 
  \subsubsection{FPA17017}
 
  

\begin{figure}[!t]
    \centering
    \includegraphics[width=12cm,height=8cm,scale=.75]{{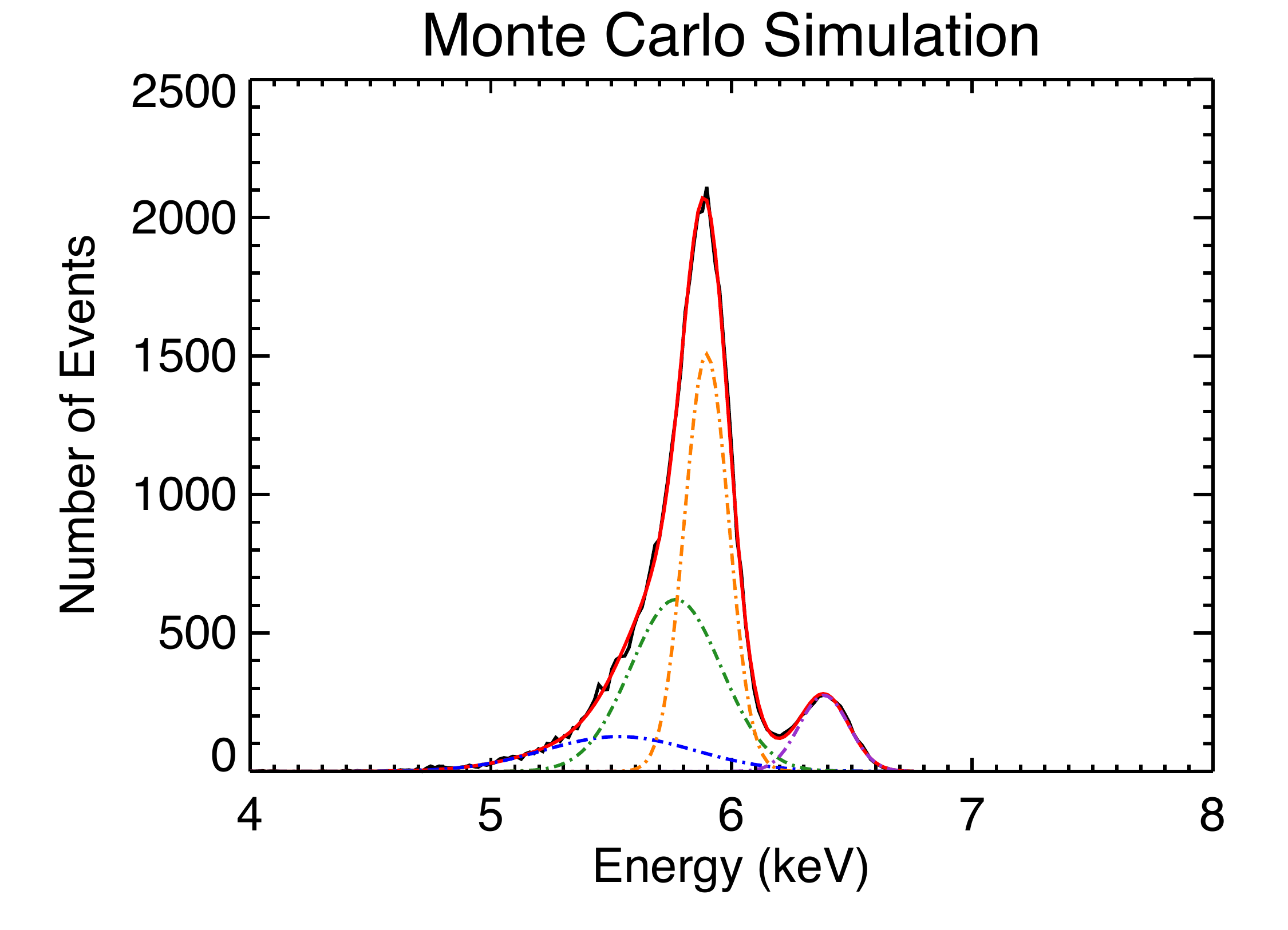}}
    \caption{FPA17017 Monte Carlo simulated spectrum using the measured gain variation map and the increased noise from the vacuum pumps, $\sigma_{RN+Pump}$.  Measured energy resolution for Mn K$\alpha$ $\Delta$E/E = 3.8 $\pm$ 0.2 \% after fitting four Gaussians to the spectrum (dashed lines) and $\Delta$E/E = 4.6 $\pm$ 0.2 \% when measuring the width at half max.  This four Gaussian fit was needed to properly fit the modeled spectrum ($\chi^{2}_{Reduced}$ $<$ 1).}
    \label{Ch6-figure:FPA17017_monte_carlo}
\end{figure}

For FPA17017, the theoretical limit from Equation \ref{e_res_theo_limit}, including the measured increased noise, $\sigma_{RN+Pump}$, of 17.8 e$^{-}$ in Gain 0 mode, is $\Delta$E/E = 3.2 \%.   To model the gain variation on FPA17017, a gain variation map with the same non-Gaussian distribution (Figure \ref{Ch6-figure:FPA17017_gain_var_hist}) was created for the Monte Carlo simulation.  A Gaussian distribution of X-ray events was created with the mean equal to the theoretical Mn K$\alpha$ and K$\beta$ mean and the standard deviation equal to the combination, in quadrature, of the deviations from the Fano factor term and the measured $\sigma_{RN+Pump}$.    The Gaussian distribution of events was convolved through the gain variation map to create a modeled spectrum. The modeled spectrum, using the increased noise, and the corresponding four Gaussian fit can be seen in Figure \ref{Ch6-figure:FPA17017_monte_carlo}.  Similar to the measured spectrum in Figure \ref{Ch6-figure:FPA17017_gain0_grade0_spec} (as described in Section \ref{energy_res}), four Gaussians were needed to fit the modeled spectrum correctly due to the low energy tail.   This provided a phenomenological fit to properly calculate the width of the spectrum around the line we were intending to measure.  The Mn K$\alpha$ line (dot-dashed Gaussian with mean at $\sim$5.89 keV in Figure \ref{Ch6-figure:FPA17017_monte_carlo}) was modeled to be $\Delta$E/E = 3.8 $\pm$ 0.2 \%, which is marginally below the four Gaussian fit measured value of $\Delta$E/E = 4.4 $\pm$ 0.2 \% with noise equal to $\sigma_{RN+Pump}$.  By simply measuring the full width at half max, the modeled spectrum has a $\Delta$E/E = 4.6 $\pm$ 0.20 \%.   The measured value is $\Delta$E/E = 6.1 $\pm$ 0.2 \% with noise equal to $\sigma_{RN+Pump}$.
 
 
 Applying the above simulation to the previous measurements, without the increase in noise from the vacuum pumps (see Section \ref{read_noise}), we modeled a $\Delta$E/E = 3.2 $\pm$ 0.2 \% using the measured read noise of 12.3 e$^{-}$ and a four Gaussian fit.  The measured value is $\Delta$E/E = 4.0 $\pm$ 0.2 \% as reported in Section \ref{energy_res}.  By simply measuring the full width of the large Mn K$\alpha$ peak at half max, we measured $\Delta$E/E = 4.0 $\pm$ 0.2 \%.  The measured value is 5.2 $\pm$ 0.2 \% as reported in Section \ref{energy_res}.  
 
Although the modeled spectra predict slightly lower values for $\Delta$E/E relative to these measured values, they are in rough agreement.  We note the existence of an extended low energy tail in the measured spectrum (Figure \ref{Ch6-figure:FPA17017_gain0_grade0_spec}), which is absent from the modeled spectrum (Figure \ref{Ch6-figure:FPA17017_monte_carlo}).   These events are likely to be caused by contamination of the Grade 0 event sample with actual Grade 1-4 events, as well as the possibility of charge loss on the detector when an event is near a bad pixel.  Events near bad pixels can be characterized as Grade 0 events, when they are in fact Grade 1-4 events.  These two factors account for some of the discrepancy seen in the measured $\Delta$E/E on the measured spectrum versus the modeled spectrum.  

The measured energy resolution for  FPA17017 is degraded compared to FPA17014, as shown in Section \ref{energy_res}, despite the detectors having similar read noise. The gain variation has been shown to account for the degraded energy resolution on FPA17017.  The modeled energy resolution, including gain variation, on FPA17014 has been shown to be consistent with the measured spectrum.  For FPA17017, the modeled value for $\Delta$E/E is in rough agreement with the measured value, but it appears the measurements of $\Delta$E/E include some small contribution beyond those that were modeled.  Other degrading factors, such as contamination of Grade 0 events and charge loss, were not modeled on this engineering grade detector.  

\section{Conclusion}
We have shown the the Speedster-EXD can successfully take images in a mode that allows the system to read out only the pixels that contain charge above a set threshold, thus allowing it to achieve faster effective frame rates. The detector has been operated in both a Full Frame Readout mode and a ``Sparse" event driven readout mode, and its characteristics as an X-ray detector have been measured in each of these modes.  The use of the CTIA amplifier has also been shown to reduce IPC to the point at which its impact is unmeasurable.  We measure the energy resolution at 5.89 keV ($\Delta$E/E = 3.5 \%) and 1.49 keV ($\Delta$E/E = 10 \%), the best measurements to date for a hybrid CMOS X-ray detector. The full characterization of the gain variation has been performed on both Speedster detectors, and a Monte Carlo simulation has been applied to better characterize the contributions to the energy resolution.  The Speedster-EXD detector has demonstrated the capability of the in-pixel circuitry of hybrid CMOS X-ray detectors and has proven to be an excellent step forward in the development of HCDs for future X-ray space missions.   


\acknowledgments 
We gratefully acknowledge support from that NASA APRA Detector Development program, particularly support from grant number NNX11AF98G.  We would also like to acknowledge Vincent Douence for helpful discussions and efforts regarding the operation and optimization of the Speedster device.







\end{spacing}
\end{document}